\documentclass[acmsmall]{acmart}

\AtBeginDocument{%
  \providecommand\BibTeX{{%
    \normalfont B\kern-0.5em{\scshape i\kern-0.25em b}\kern-0.8em\TeX}}}

\usepackage{graphicx}
\usepackage{algorithmic}
\usepackage{listings}
\usepackage{hyperref}

\graphicspath{ {./graphs/} }

\newcommand{\R}{\mathbb{R}}

\begin{document}

\title{ACORNS: An Easy-To-Use Code Generator for Gradients and Hessians}

\author{Deshana Desai}
\authornote{Both authors contributed equally to this research.}

\affiliation{%
  \institution{New York University}
}

\author{Etai Shuchatowitz}
\authornotemark[1]

\affiliation{%
  \institution{New York University}
}

\author{Zhongshi Jiang}

\affiliation{%
  \institution{New York University}
}

\author{Teseo Schneider}
\affiliation{%
  \institution{New York University}
}

\author{Daniele Panozzo}

\affiliation{%
  \institution{New York University}
}


\begin{abstract}
% !TeX root = ./main.tex

The computation of first and second-order derivatives is a staple in many computing applications, ranging from machine learning to scientific computing. We propose an algorithm to automatically differentiate algorithms written in a subset of C99 code and its efficient implementation as a Python script. We demonstrate that our algorithm enables automatic, reliable, and efficient differentiation of common algorithms used in physical simulation and geometry processing.
\end{abstract}



% \keywords{datasets, neural networks, gaze detection, text tagging}

\maketitle

% !TeX root = ./main.tex

\section{Introduction}

The calculation of derivatives is used in many applications, including machine learning, scientific computing, geometry processing, computer vision, natural language processing, and many more. The calculation of derivatives of large and complex computational graphs is a computational bottleneck for these applications. Thus the efficient computation of derivatives is an area of active research interest. Extensive research and engineering efforts have been spent in the last years to support applications in deep learning, that led to the development of libraries such as TensorFlow or PyTorch, which target applications with large and dense tensors. The availability of easy-to-use differentiation libraries is one of the reasons for the massive advancement in deep learning. They allow researchers to focus on the design of new network architectures and loss functions, while automating and ensuring correctness in the low-level tasks required to minimize these functionals. However, they are specialized for computing first derivatives only, and they are optimized for large, dense tensors, which are uncommon in applications outside of deep learning. The novelty in these libraries is not in the automatic differentiation technique used, but in the ease of use, cross-platform availability, and ease of integration in new research projects.

Our goal is to provide a similar solution for scientific applications requiring first and second derivatives of expressions involving small to medium tensors, using an approach based on symbolic differentiation. When paired with the optimization algorithms available in modern compilers, our approach generates code around an order of magnitude faster than existing methods while allowing users to write their functions directly in the C language. The algorithm is easy to integrate into existing build systems and produces dependency-free code.

The core parts of the library are implemented in Python using a C99 parser~\cite{benderskypycparser}. We perform symbolic differentiation with unrolled derivatives, stored as expression trees, and generate an efficient kernel to evaluate them using an existing C99 compiler, addressing the potentially long compilation times with a file-splitting approach. State-of-the-art compilers are capable of sophisticated optimizations such as loop and expression trees vectorization, elimination of redundancy in expressions, and reusing computations  performed in previous iterations of loops. Our approach benefits from these features, and the performance of our method will likely increase as compiler technology progresses. The expression unrolling also allows us to trivially parallelize computations not only for different data points but also for a single data point over multiple CPU cores, which are common in modern architectures. Finally, we provide support efficient differentiation of algorithms with nested loops, arrays, and functions. A key difference with existing libraries, is that we create kernels to efficiently evaluate the expression itself, its gradient, and its Hessian, all optionally supporting parallel evaluation.

Our open-source implementation makes the integration of the algorithm straightforward in modern C and C++ applications, thanks to its ability to directly differentiate code and to produce standard, dependency-free C99 code that can be easily used in existing applications.

To demonstrate the practical impact of our approach, we integrated it into two applications, one in geometry processing and one in scientific computing. We show that our library provides important speedups in the overall running time of mesh parametrization algorithms and in the quasi-static simulation of non-linear elastic deformations.

% !TeX root = ./main.tex

\section{Related Work}
Automatic (or algorithmic) differentiation has been introduced in the pioneering works of \cite{wengert1964simple,speelpenning1980compiling}. We briefly review the most relevant work and refer readers to various books and survey for further detail \cite{naumann2012art,griewank2008evaluating,bischof2000computing,margossian2019review} and \url{www.autodiff.org} for a collection of implementations. We provide comparisons against representative methods for operator overloading and source transformation for both gradients and higher order derivatives in Section \ref{sec:results}.

\subsection{Operator Overloading}
The most common way of implementing automatic differentiation is through defining a new object class, which associates the gradient information to the data. In forward mode, the gradient information is updated throughout the trace of the computation. And in reverse mode, a tape of computation is recorded, and the gradient information is then (back)propagated following the recorded tape. Due to the versatility and freedom of prototyping,
operator overloading based approaches have gained a lot of attraction, and notable examples include
\cite{paszke2019pytorch,paszke2017automatic,abadi2016tensorflow,bell2012cppad,griewank1996algorithm,Mitsuba}.
However, the efficiency of the code often suffers due to the additional runtime overhead, which is especially noticeable when the same computation is executed repeatedly. 

\subsection{Source Transformation}
Source transformation approaches parse the input source code (most commonly C and Fortran \cite{hascoet2013tapenade,vossbeck2008development}) and generate a new algorithm that computes the derivatives of the input function.
Compared to operator overloading approaches, obtaining the source code file \emph{a priori} opens the doors to global data flow analysis and optimization, thus reducing the running time. Additionally, this approach enables the developers to directly debug the code for derivatives computation, a crucial feature to identify and address potential issues with numerical round-off or overflows. The reliance on compilers to compile the generated source-code makes these approaches future-proof, in the sense that the generated code will benefit from progress on compiler design and optimization.

A limitation of source transformation is that, since the flow is assumed fixed to allow for the above optimization, it is challenging to support conditionals or unbounded loops. In our algorithm, we restrict to a subset of C99 covering common use-cases in scientific computing, finite element analysis, and computer graphics.

A minor, yet practically relevant, drawback of source code transformation is the presence of the additional source files, posing challenges on the build system and version control of the evolving software. We show that our library, thanks to its permissive open-source license, minimal dependencies, and availability in the Conda package system, is effortless to integrate into a CMake-based build system \cite{martin2010mastering}. We show how to automatically generate the required files as part of the build process, making the automatic differentiation transparent to the user (Section \ref{sec:build}).

\subsection{Higher Order Derivatives}
Although valuable in scientific computing and numerical optimization,
computing higher-order derivatives is challenging due to the need of memory management and sparsity pattern detection\cite{gower2010hessian,walther2008computing}. 
Therefore, only few existing libraries support second or higher-order derivatives computation.
Our method supports dense Hessian computation, taking advantage of the symmetric structure to reduce computational times. %Integration with sparsity analysis to scale up is a valuable next step 

% \subsection{Common Implementation for AD}
\section{Method}

The \emph{input} of our algorithm is an algorithm written in a subset of C99\cite{ritchie1988c}: we support arrays, for loops, nested loops, binary assignments, functions, and variable declarations.
The \emph{output} is a set of self-contained, multi-threaded C99 functions to evaluate the function in the input file, its first derivative, and its second derivative.
We designed our library to become part of a build system, seamlessly generating the derivative code as part of the build procedure of a software package (Section \ref{sec:build}).

We overview our algorithm in Section \ref{sec:overview}, and show a complete execution on a prototypical expression in Section \ref{sec:example}. We then discuss how to parallelize the evaluation in Section \ref{sec:parallel} and how to make our approach scale to large expressions in Section \ref{sec:scalability}. We show the performance of the algorithm on synthetic expressions (Section \ref{sec:results}), comparing it with state-of-the-art differentiation libraries, and integrate it into a physical simulation library for a  realistic benchmark.

\subsection{Overview}
\label{sec:overview}

\paragraph{Primer on Automatic Differentiation.} Given a target function $f \colon \R^n \to \R^m$, the corresponding Jacobian matrix has $n \times m$ entries $J_{ij} = \frac{\partial f_i}{\partial x_j}$.
We use the pyc parser \cite{pycparser} to parse the input C99 code to produce an Abstract Syntax Tree (AST) of the input algorithm. The leaves of the tree are either numerical constants or variables and the internal nodes follow a hierarchical structure built from the code.
%The AST is parsed to locate the sub-tree containing the declaration of the target function $f$.
Let us consider an example, with $f$ a composite function: $f(x) = h \circ g(x) = h(g(x))$, with $x \in \R^n$, $g\colon \R^n \to \R^m$ with
\begin{equation*}
    J_{ij} = \frac{\partial f_i}{\partial x_j} = \frac{\partial h_i}{\partial g_1} \frac{\partial g_1}{\partial x_j} + \frac{\partial h_i}{\partial g_2} \frac{\partial g_2}{\partial x_j} + ... + \frac{\partial h_i}{\partial g_k} \frac{\partial g_k}{\partial x_j}.
\end{equation*}

% More generally, if our target function $f$ is the composite expression of $L$ functions $f = f^L\;o\;f^{L-1}\;o\;...\;o\;f^1$ then the corresponding Jacobian matrix verifies:
% \begin{equation}
%     J = J_L. J_{L-1} ... J_1
% \end{equation}

From the example we see that the auto-differentiation of a function $f$ breaks down the action of the Jacobian matrix on a vector into simple components, which are evaluated sequentially. We store the operations required for the calculation of the derivative of each of these pieces as a string. The final derivative is the accumulation of the piece-wise derivatives of the sub-operations following  the chain rule. %We can execute this method for both forward and reverse differentiation methods however the final derivative has equal complexity.
% Higher order derivatives are computed recursively: To compute the $n^{th}$ derivative, the expression tree containing the target function is differentiated n times.\DP{Do we do more than 2 order? If not drop.}

%The computed derivative formulas are written as C code to a file.
% \begin{equation}
%     \frac{\delta f^n(x)}{\delta x^n} = \frac{\delta f^{n-1}}{\delta x^{n-1}} \left( \frac{\delta f(x)}{\delta x} \right)
% \end{equation}

For computing second order derivatives, we differentiate the tree twice. We further optimize over the number of calculations by taking advantage of the symmetry of the Hessian: we compute only the lower triangular part of the matrix and copy it over the upper triangular half.

% We compute derivative formulas for the upper triangular matrix of the n x n matrix and simply assigning the values for the remaining matrix entries to the above.

% \begin{equation}
%     \delta_{x_1,x_2} f = \delta_{x_2,x_1} f
% \end{equation}
%\DP{this requires an example to clarify}

\paragraph{Implementation Specific Details.}
The derivative formulas are unrolled in a sequence of assignments (no loops or conditional are used) and saved on a C file. The file is then compiled with full code optimization (\texttt{-O3}  with \texttt{gcc} \cite{stallman2002gnu} and \texttt{clang} \cite{lattner2004llvm}). This leverages the compiler's ability to perform redundancy elimination on trees, reusing computations (especially memory loads and stores), and automatically vectorizing the code when appropriate.
% Reference: https://gcc.gnu.org/onlinedocs/gcc-4.4.7/gcc/Optimize-Options.html

% The source function $f$ is written in a subset of C99. We support the following functionalities: nested loops, conditional statements, variable assignments, different function declarations and multi-dimensional arrays as data structures. \TS{repeated}
In the case of loops or conditional statements being present in the program, we perform an additional parsing step to expand these structures. We perform loop unrolling to produce a sequence of operations and eliminate other instructions that control the loop. Conditional statements are removed by evaluating their condition during parsing. The output of this phase is a list of assignment instructions, which are stored in an intermediate binary format. This file is then parsed to create the AST, differentiated, and the differentiated AST is exported as a C file. %This additional step of parsing is required to unroll the loops and create a computational graph from the programming structures. This step increases redundancy, however, the motivation is to leverage the compiler's abilities for redundancy elimination.

\subsection{Example}
\label{sec:example}
\begin{figure}
    \centering
    \includegraphics[width=0.5\linewidth]{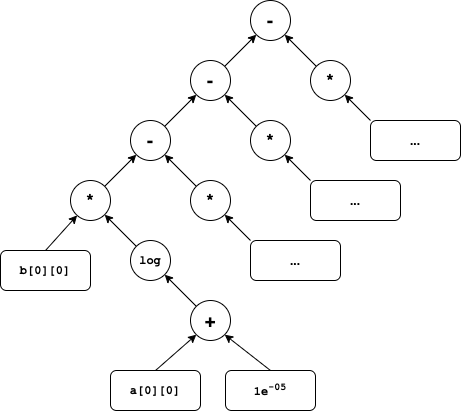}
    \caption{Expression graph for \emph{cross entropy} loss function}
    \label{fig:ast}
\end{figure}

To further illustrate the mechanisms of ACORNS, we consider a typical example from statistics. We will compute the gradient of the \emph{cross entropy} loss function, between two probability distributions $a$ and $b$. The loss function can be given by the following piece of C99 code:
\begin{lstlisting}
double cross_entropy(const double **a, const double **b){
    double loss = 0;
    for(int i = 0; i < 2; i++){
        for(int j = 0; j < 2; j++ ){
            loss = loss - (b[i][j] * log(a[i][j] + 0.00001));
        }
    }
    return loss;
}
\end{lstlisting}
The function is unrolled to produce the following code:

\begin{lstlisting}
double cross_entropy(const double **a, const double **b){
    double loss = 0;
    loss = (loss) - ((b[0][0]) * (log((a[0][0]) + (0.00001))));
    loss = (loss) - ((b[0][1]) * (log((a[0][1]) + (0.00001))));
    loss = (loss) - ((b[1][0]) * (log((a[1][0]) + (0.00001))));
    loss = (loss) - ((b[1][1]) * (log((a[1][1]) + (0.00001))));
    return loss;
}
\end{lstlisting}
An {Abstract Syntax Tree} (Figure \ref{fig:ast}) is constructed from the unrolled code.
The syntax tree is differentiated using back-propagation to construct the derivative equations. This unrolled code is finally compiled by a C compiler with full code optimization enabled.
\begin{lstlisting}
for(int p = 0; p < num_points; ++p)
{
    ders[i*2+0]+= ((((((((0) - (((log((a[0][0] + 0.00001))) * (b[0][0]) + b[0][0] *
    ((1/((a[0][0] + 0.00001))*0)))))) - (((log((a[0][1] + 0.00001))) * (b[0][1]) +
    b[0][1] * ((1/((a[0][1] + 0.00001))*0)))))) - (((log((a[1][0] + ...
}
\end{lstlisting}

\subsection{Parallelization}
\label{sec:parallel}
%\DP{Explain why it is trivial  for us to parallelize}

% The execution time and memory requirement, especially for computing second order derivatives,  of the original program increases by $n^2$, relative to the original program. Especially for large n, the execution time of the higher order derivatives dramatically increases. To overcome this situation, we suggest the parallel execution of the differentiated code.

To accelerate the evaluation of large gradients or Hessians, we observe that the evaluation of the different entries is independent, and can thus be trivially parallelized using OpenMP\cite{chandra2001parallel} directives. We experimentally study the scaling of our generated code in Section \ref{sec:results}.
% We suggest the following strategies for automatically parallelizing the generated source code:

% \begin{enumerate}
%     \item For gradients, the entire differentiated code is executed in parallel over different evaluation points. The derivative computation for a single data point remains serial. \DP{why not parallelize this too for a single evaluation?}
%     \item For Hessians, we divide the entry computations into parallel portions such that every thread has a balanced number of entries to compute. \DP{we need to say when we do it, only for large ones I imagine?}
% \end{enumerate}

% As an example, we demonstrate the the computation of second order derivatives of a simple expression using the parallelization strategies described above.

% \begin{algorithmic}
%     FUNCTION forward_diff(N, ders, a, b, c)
%     openmp parallel for i=1:N
%           ders[i*0+0] = a*b+c*b;
%     openmp parallel for end
% \end{algorithmic}

% \begin{algorithmic}
%     FUNCTION forward_diff(N, ders, a, b, c)
%     openmp parallel for i=1:N
%           openmp parallel section
%           ders[i*0+0] = a*b+c*b;
%           ders[i*0+1] = a*b+c*b;
%           openmp parallel section
%           ders[i*0+0] = a*b+c*b;
%           ders[i*0+1] = a*b+c*b;
%           openmp parallel section end
%     openmp parallel for end
% \end{algorithmic}

\begin{figure}
\parbox{0.02\linewidth}{\centering\rotatebox{90}{\scriptsize{Time (s)}}}\hfill\hfill
\parbox{.32\linewidth}{\centering
    \includegraphics[width=\linewidth]{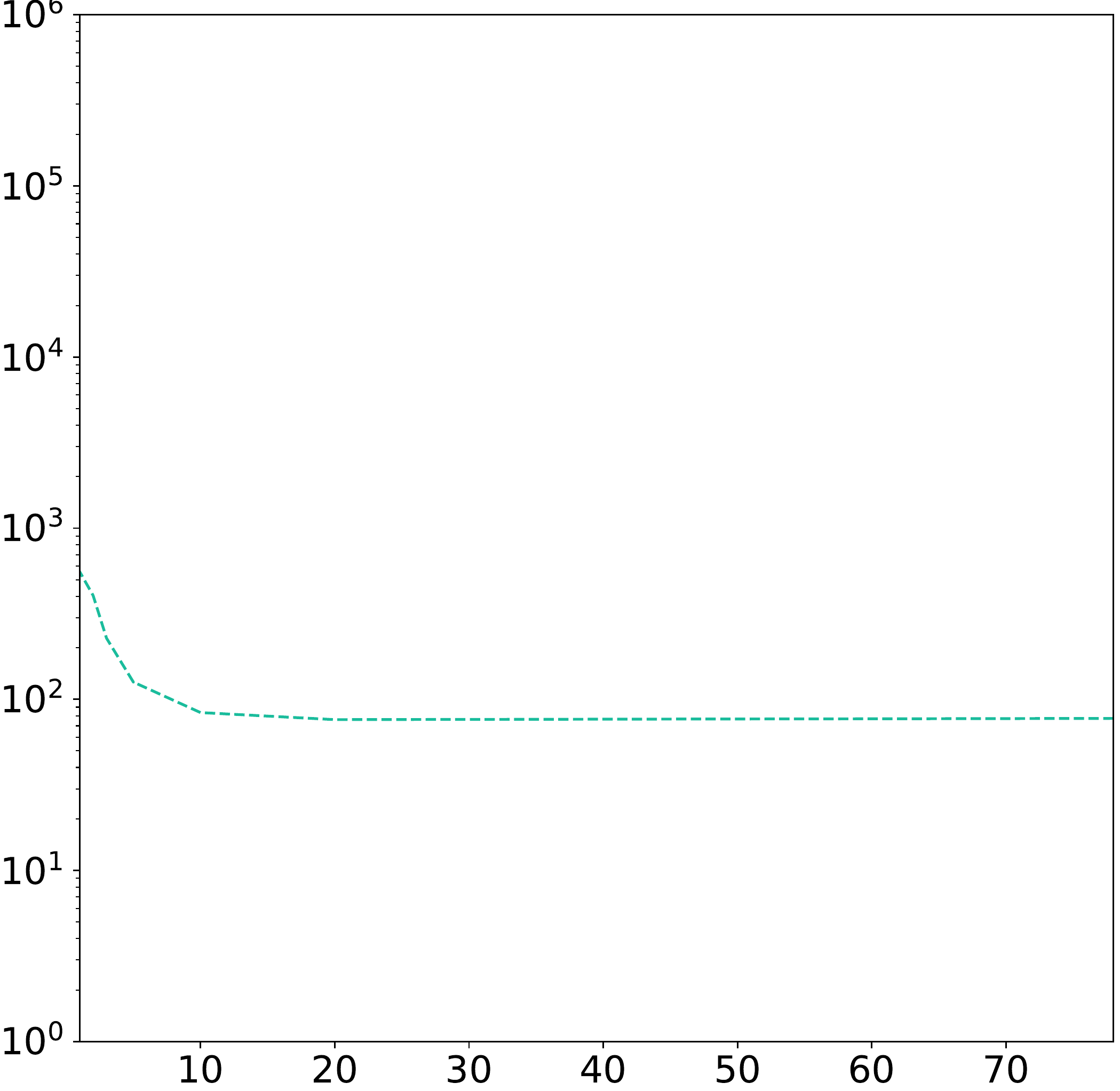}
    \scriptsize{Number of Files}
}
\parbox{.32\linewidth}{\centering
    \includegraphics[width=\linewidth]{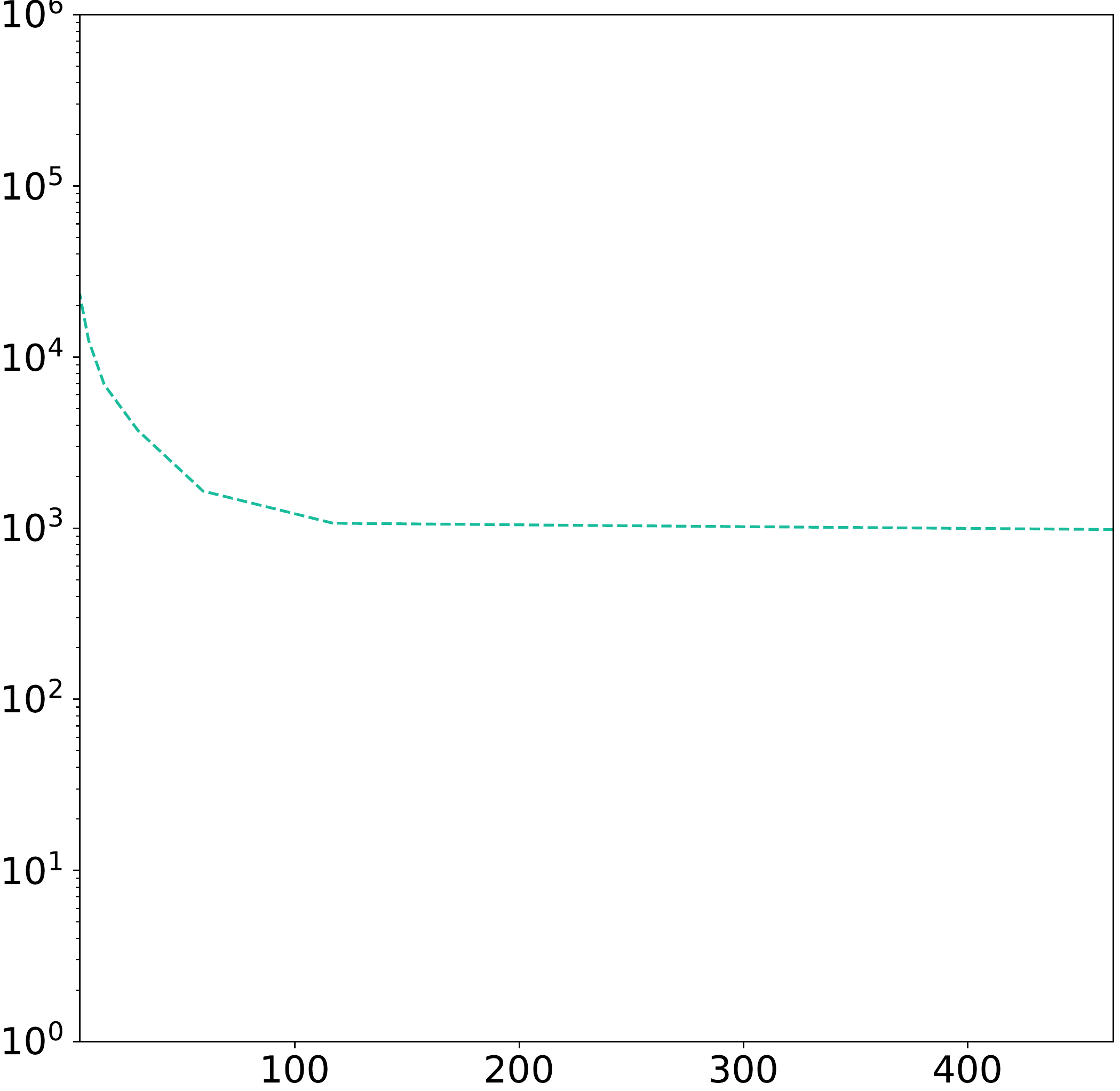}
    \scriptsize{Number of Files}
}
\parbox{.32\linewidth}{\centering
    \includegraphics[width=\linewidth]{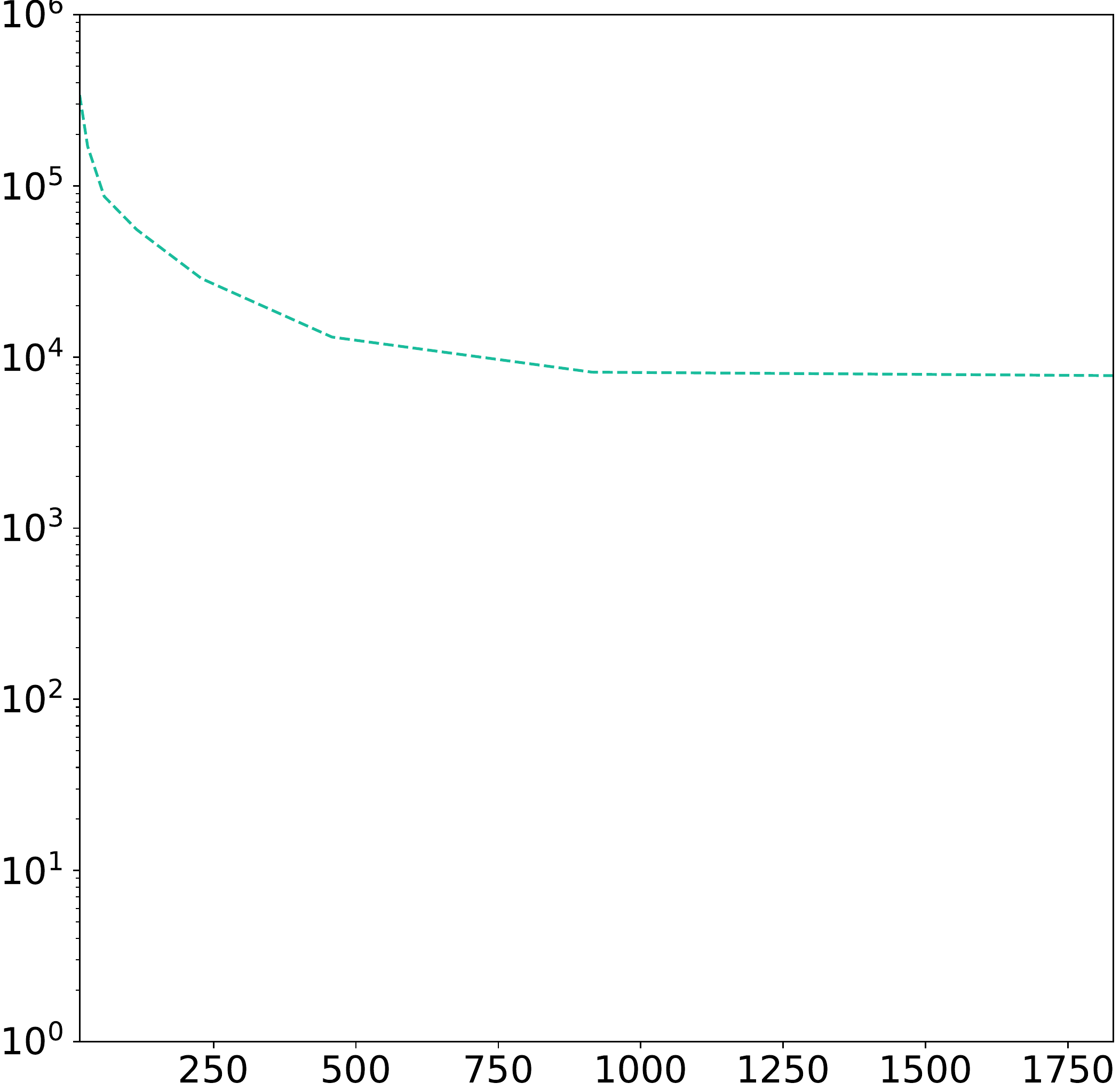}
    \scriptsize{Number of Files}
}
\caption{Compile times for Hessians of a function of 12, 30, and 60 parameters.}
\label{fig:compile-times-hessian}
\end{figure}

\subsection{Large Expressions}
\label{sec:scalability}
As the complexity of the expressions and/or the number of outputs grows, the size of the generated C file grows accordingly. While this opens more opportunities for the compilers to optimize the evaluation, it also increases memory consumption and compilation time. For large expressions, such as the Hessian of a cubic Lagrangian element, this might become impractical.
We propose a simple, yet effective and practical, strategy to overcome this issue: we split the output into multiple C files that are compiled independently. This has the downside of reducing the optimization opportunities of compilers. However, in our experiments, the performance drop is negligible and we thus use this option aggressively to reduce compilation times.

% The compilation time also becomes unwieldy for large programs as the hessian computations grow by $O(n^2)$. This also places a large burden on the register allocator. Different compilers can have different capabilities in handling optimizations for redundancy eliminations on large files. Upon successful optimizations, the file size of the compiled program is drastically smaller than the source code file. To overcome large compilation times, we explore splitting the large file into smaller files which can be compiled separately. This would be accompanied by a trade-off in optimizations performed by the compiler since the optimizations would now be made locally for the smaller files as opposed to over the entire file.

% We thus split the computation of the $n(n+1)/2$ entries of the hessian over k files. The ordering of the entries are recorded during the split. At runtime, we re-order and concatenate the k arrays into a single one.

%\subsection{Evaluation}

To evaluate the effect of the split on performances, we consider a function for calculating the Neo-Hookean Elasticity energy \cite{rivlin1948large} for a linear, quadratic, and cubic Lagrange element (12, 30, and 60 variables respectively). The code generated by these function is large (29 GB for the function with 60 variables) and the compilation with gcc9 does not terminate in 160 hours.

We tested splitting into different number of files, measuring  compilation time, evaluation time, and binary size after compilation.

\begin{figure}
\parbox{0.02\linewidth}{\centering\rotatebox{90}{\scriptsize{Size (Mb)}}}\hfill\hfill
\parbox{.32\linewidth}{\centering
    \includegraphics[width=\linewidth]{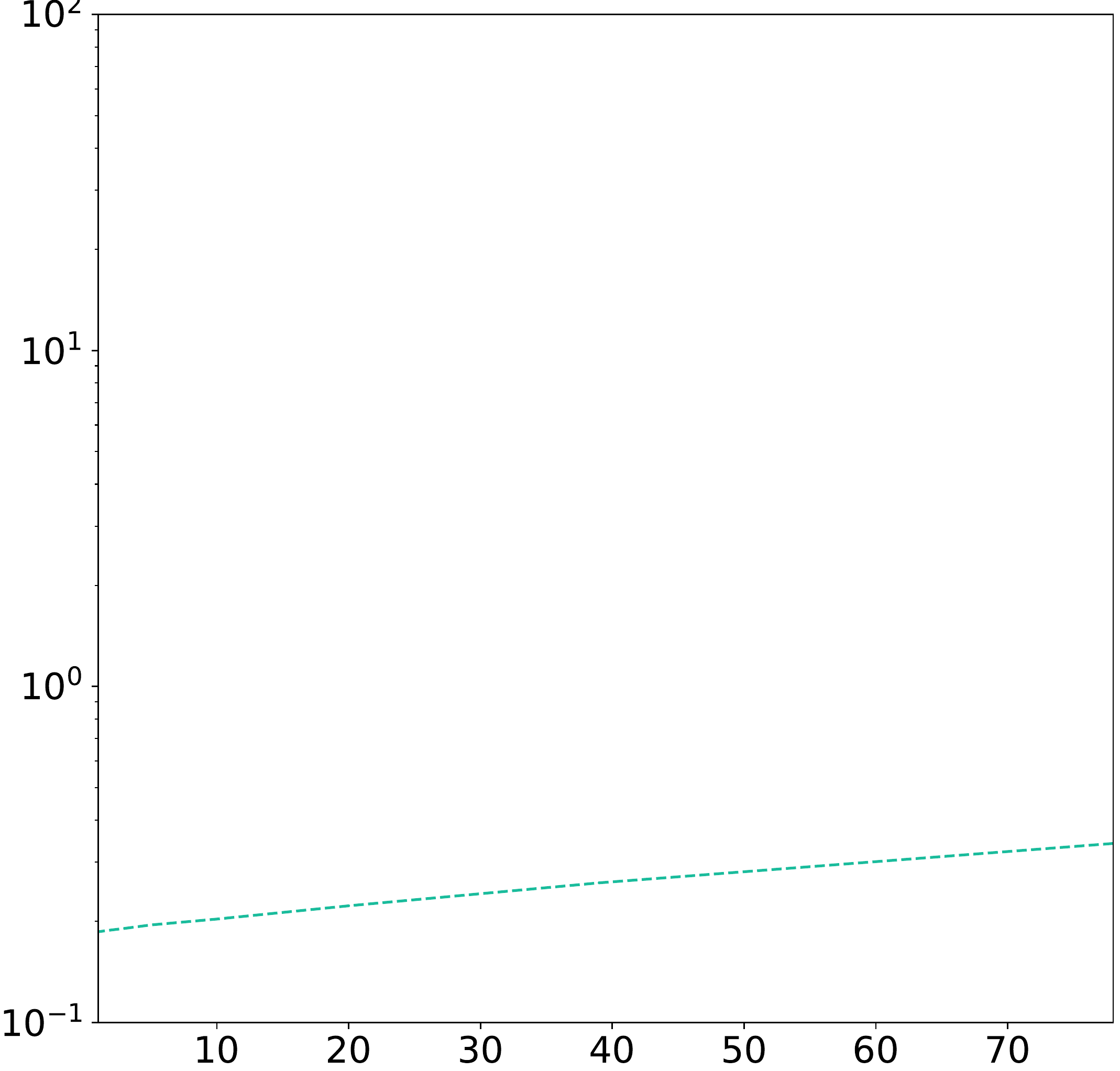}
    \scriptsize{Number of Files}
}
\parbox{.32\linewidth}{\centering
    \includegraphics[width=\linewidth]{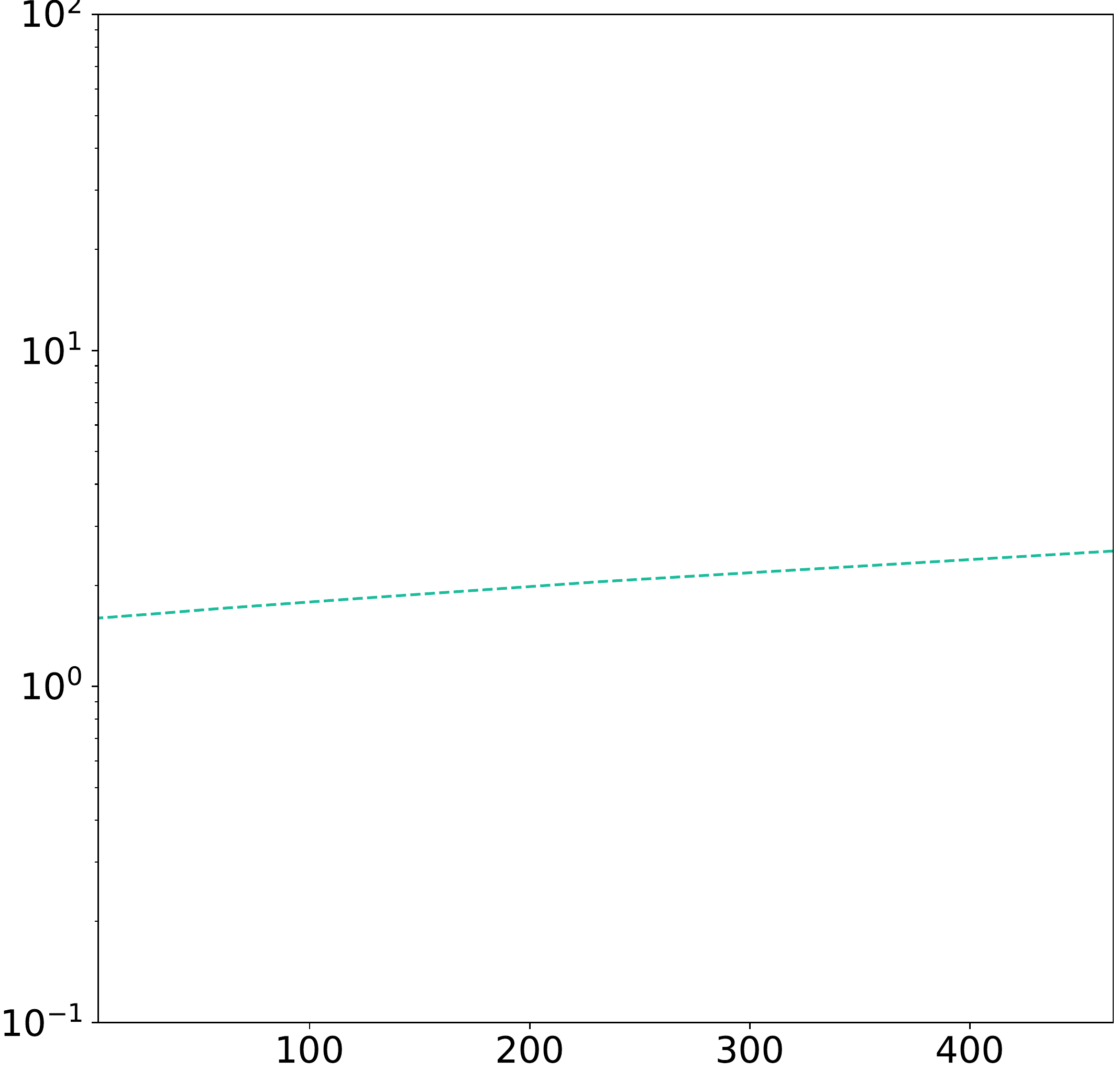}
    \scriptsize{Number of Files}
}
\parbox{.32\linewidth}{\centering
    \includegraphics[width=\linewidth]{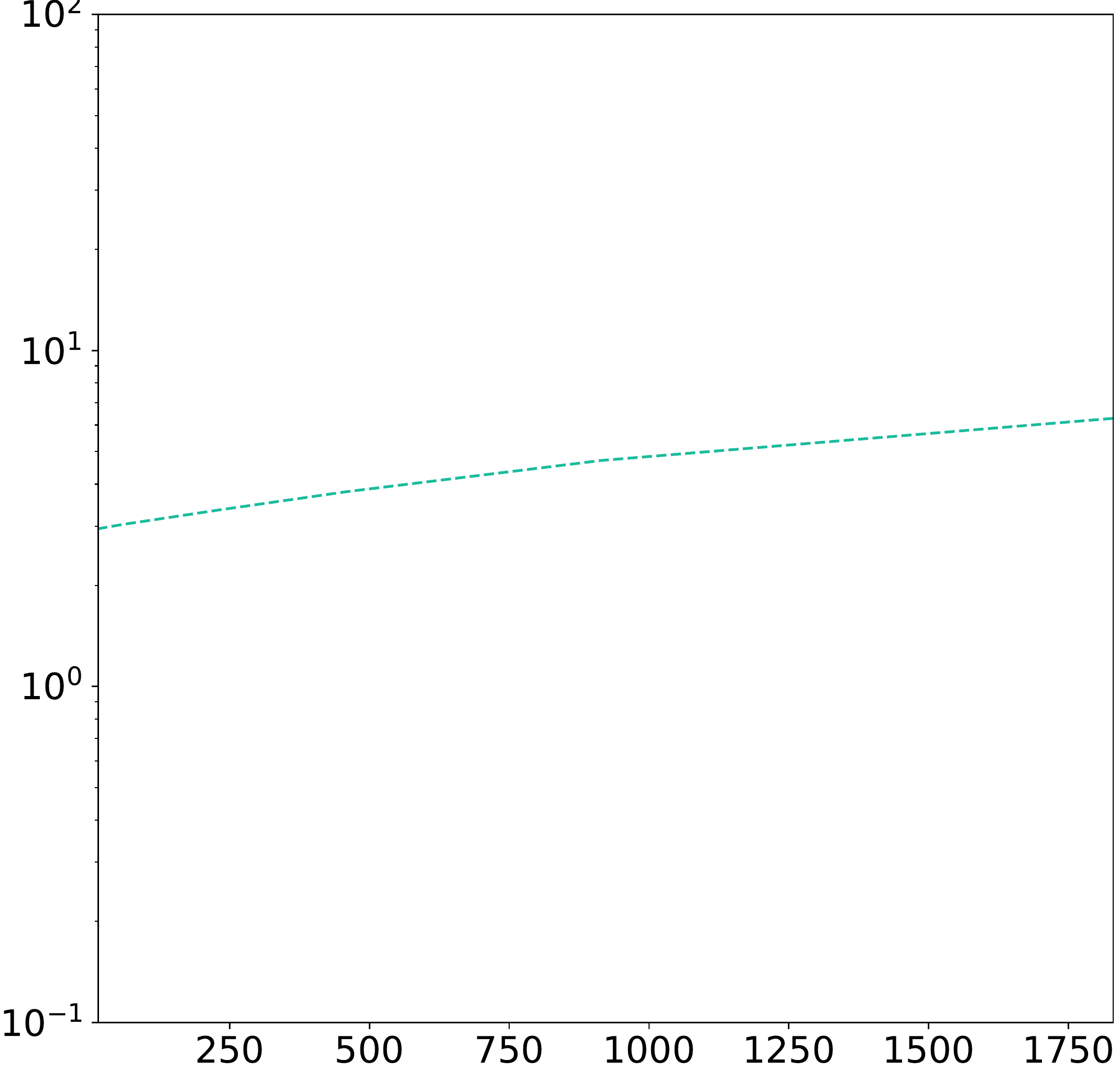}
    \scriptsize{Number of Files}
}
\caption{Size of the generated binary files (.o) for Hessians of a function of 12, 30, and 60 variables.}
\label{fig:o-sizes-polyfem}
\end{figure}

\begin{figure}
\parbox{0.02\linewidth}{\centering\rotatebox{90}{\scriptsize{Time (s)}}}\hfill\hfill
\parbox{.32\linewidth}{\centering
    \includegraphics[width=\linewidth]{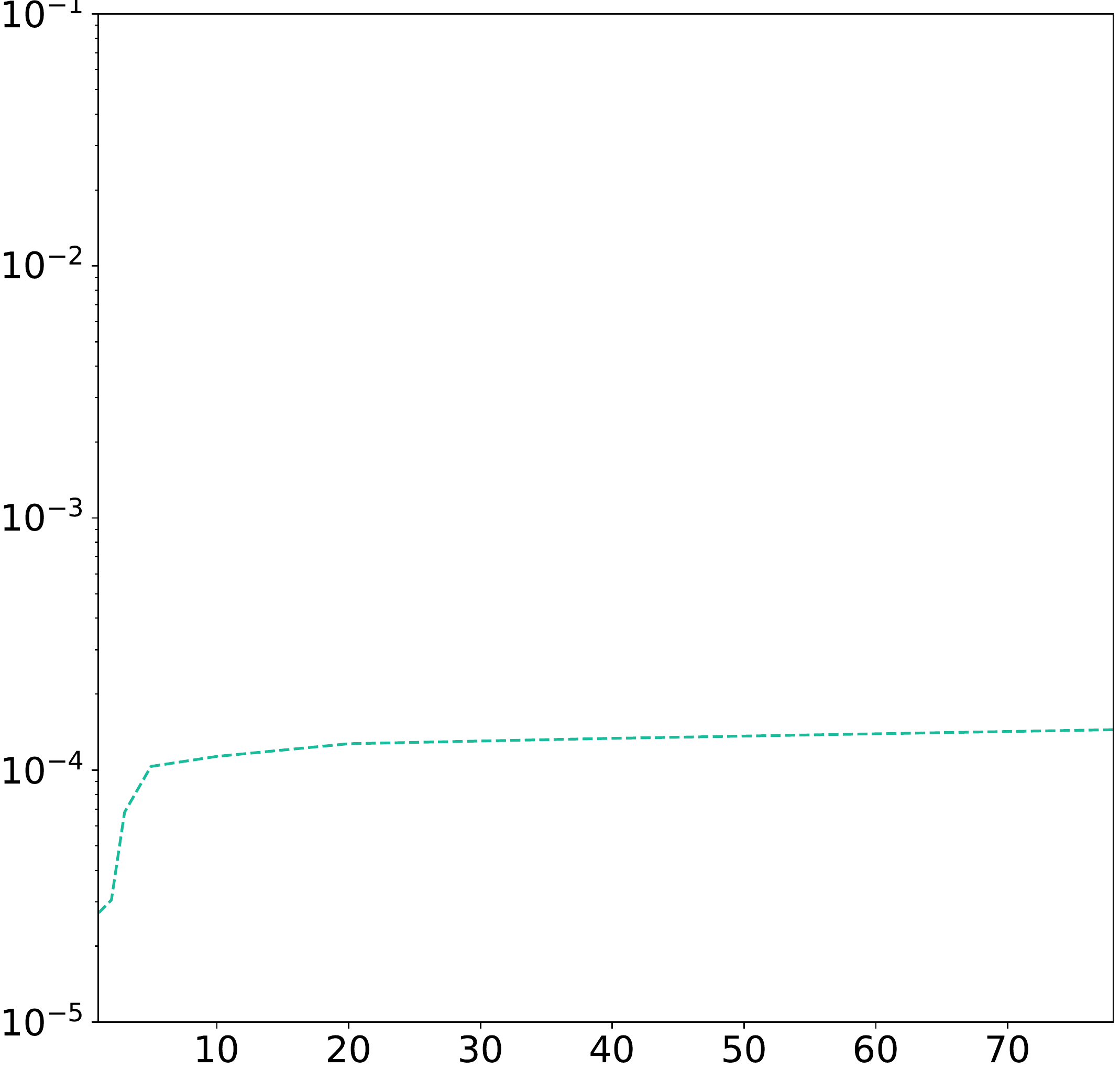}
    \scriptsize{Number of Files}
}
\parbox{.32\linewidth}{\centering
    \includegraphics[width=\linewidth]{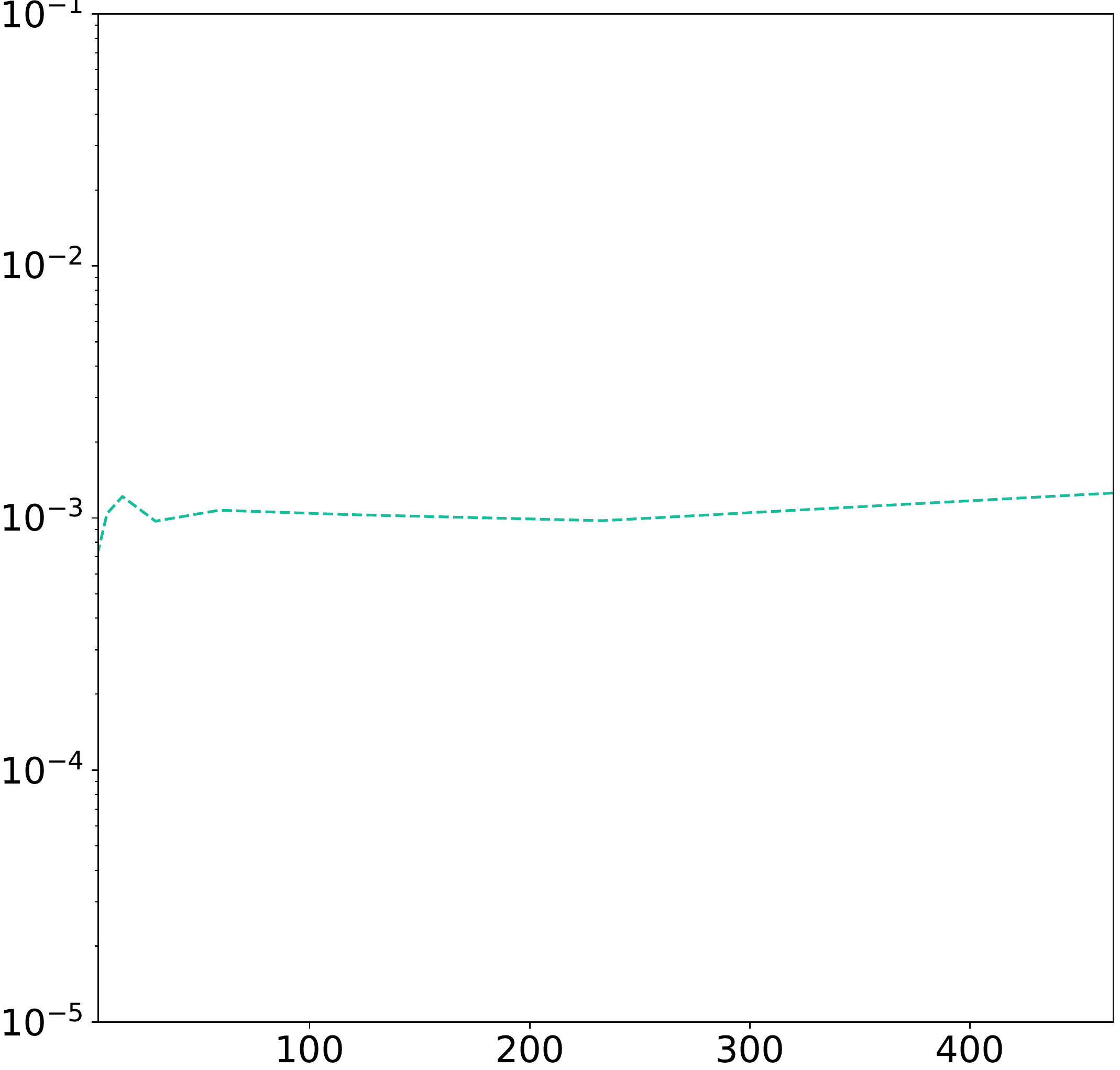}
    \scriptsize{Number of Files}
}
\parbox{.32\linewidth}{\centering
    \includegraphics[width=\linewidth]{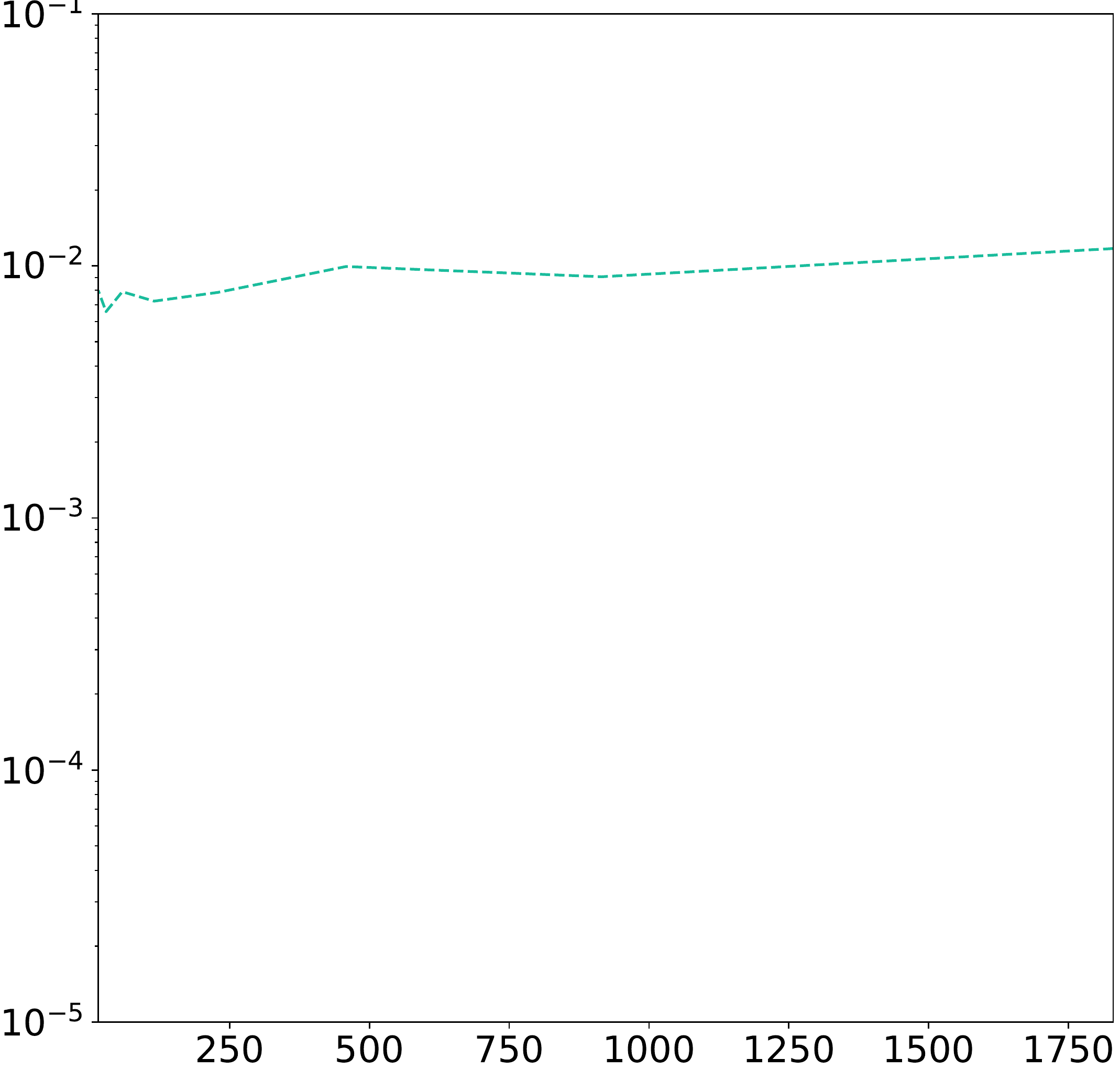}
    \scriptsize{Number of Files}
}
\caption{Run times for Hessians of a function of 12, 30, and 60 variables.}
\label{fig:runtimes-polyfem}
\end{figure}

As expected, the compilation time decreases as the files become smaller (Figure \ref{fig:compile-times-hessian}) while the size of the binary generated is not noticeably affected by the split (Figure \ref{fig:o-sizes-polyfem}). Surprisingly, the evaluation time of the generated code is also not noticeably affected by the split (Figure \ref{fig:runtimes-polyfem}). As a default, we thus opt for splitting so that each file had a size of around 16 MB, which is a good compromise between number of files and compilation time.

% \DP{we need to see the new data and discuss, this does not make much sense to me.}
% So, we wanted to see how file size, binary size and compilation times vary with respect to x. Looking at \ref{fig:c-sizes-polyfem} we see how file size doesn't vary with splitting things up. However, files can get very large. Even with just 60 variables we end up with C files that are on the order of 10GB. However, looking at \ref{} the final size of the binaries is only 6MB.

% However, when we applied this to a function of 60 variables, this became very unwieldy and trying to compile anything with large split size took a very long time (see Figure \ref{}). That being said, with a split-size=1 (1830 files) we were able to compile in 129.71 minutes. For contrast, with a split-size=128 (15 files) it took 94.11 hours. It's our recommendation then, that if many files doesn't matter, to keep all files as small as possible to allow for speedy compilation.

% The other thing to analyze is how long it takes to generate the files. As before we do much much better with many files, each with one array value in it. As we scale, file generation time can get very unwieldy as you can see in Figure~\ref{fig:gen-time-hessian}.\DP{why? the cost should be identical, why is it increasing?}

% \begin{figure}
% \includegraphics[width=.3\linewidth]{graphs/3D_P1_non_zero-Gen_Times.pdf}\hfill
% \includegraphics[width=.3\linewidth]{graphs/3D_P2_zero-Gen_Times.pdf}\hfill
% \includegraphics[width=.3\linewidth]{graphs/3D_P3_non_zero-Gen_Times.pdf}\hfill
% \caption{File Generation Times of files for Hessians of 12, 30, and 60 variables. \DP{Same y range} \DP{this does not make sense}}
% \label{fig:gen-time-hessian}
% \end{figure}

\subsection{Integration in a CMake Build System}
\label{sec:build}

One of the key design goal of our system is the easy and fast deployment in numerical based methods. We show how ACORNS can be integrated within a CMake build system for quick and easy deployment. In the following we describe the process and we refer to our github repository for the complete example \url{https://github.com/deshanadesai/acorns}. Let us consider a simple problem: we want to minimize the following function (saved in the file \texttt{function.c}) using gradient descent.
\begin{lstlisting}
int function_0(double x){
    double energy = pow(x, 4) - 3*pow(x, 3) + 2;
    return 0;
}
\end{lstlisting}
To use ACORNS to its derivative, we need to add the code generation to the CMake project file:
\begin{lstlisting}
cmake_minimum_required(VERSION 3.1)
project(SimpleProject)

file(MAKE_DIRECTORY ${PROJECT_SOURCE_DIR}/ders)
add_custom_command(
OUTPUT
    ${PROJECT_SOURCE_DIR}/ders/der_0.c
    ${PROJECT_SOURCE_DIR}/ders/der_0.h
COMMAND
    acorns_autodiff
        ${PROJECT_SOURCE_DIR}/functions/function_0.c energy
        --vars x --func function_0
        --output_filename ${PROJECT_SOURCE_DIR}/ders/der_0
DEPENDS
    ${PROJECT_SOURCE_DIR}/functions/function_0.c
)
add_library(derivative "ders/der_0.c")
target_include_directories(derivative PUBLIC ders)

add_executable(main_target src/example.c)

# Link target derivative to main_executable
target_link_libraries(main_target PUBLIC derivative)
\end{lstlisting}
This will generate the \texttt{ders\_0.h} and \texttt{ders\_0.c} file, which will be directly compiled by CMake and linked with the other files in the project. To use the derivatives it is then sufficient to include the generated header file. We show an example of using the computed derivative to find the root by using gradient descent.
\begin{lstlisting}
#include <der_0.h>
#include <stdio.h>
#include <math.h>

int main(int argc, char** argv) {
    double vals[1] = {6.0};
    double ders[1] = {0.0};

    int iteration;
    for (iteration = 0; iteration < 10000; iteration++) {
        // calculate derivative using ACORNS
        compute(vals, 1, ders);

        float step = 0.01 * ders[0];
        vals[0] -= step;

        if (fabs(step) <= 1e-5)
            break;
    }

    printf("Minimum: %f\n", vals[0]);
    printf("Iterations: %d\n", iteration);
    return 0;
}
\end{lstlisting}

% \TS{repeated}
% With our system, it is possible to compute automatically and efficiently gradient and Hessian of complex functions in a way that is transparent for the developer.

\section{Results}\label{sec:results}

Our algorithm is implemented in Python. We run our experiments on a 2.35 GHz AMD EPYC\texttrademark~7452 running Ubuntu 19.10 GNU/Linux 5.3.0-29-generic x86\_64. The reference implementation of ACORNS is freely available and can be easily obtained trough a conda-forge package\footnote{It can be installed with \texttt{conda install acorns -c conda-forge }.} under MIT license. The source code and the scripts to generate the experiments are available at \url{https://github.com/deshanadesai/acorns}.

We compare our generated code against 4 state-of-the-art differentiation algorithms:
\begin{enumerate}
\item{Tapenade}~\cite{hascoet2013tapenade} shares many similarities with our system. It converts C code into a C program computing its derivatives and Hessians\footnote{Hessian computation in Tapenade is not officially supported, we used the semi-manual construction proposed in the Q\&A section of the documentation for the comparison.}.

\item{PyTorch}~\cite{paszke2019pytorch} is a backward auto-differentation library tailored for machine learning. It provides both a CPU and a GPU version, we compare against the CPU version only.

\item{Mitsuba Autodiff} \cite{Mitsuba,grinspun2003discrete} compute automatic differentiation using operator overloading and depends on Eigen~\cite{code:eigen} for the linear algebra and storage. We run all our comparisons using only static stack-allocated matrices, since the dynamic mode is slower and unnecessary for the expressions we use for benchmarking.

\item{Enoki}~\cite{enoki} is a new library developed specifically to support differentiable rendering. Similarly to PyTorch, it supports both CPU and GPU evaluation, and we compare only with the CPU version and only for gradients, since it does not support Hessian computation.

\end{enumerate}

\paragraph{Expression Types}\label{sec:expression-types}
We use three expression to evaluate different realistic scenarios where autodifferentiation is used: 
\begin{enumerate}
    \item An high-order polynomial, a non-polynomial function, and procedural expressions with large number of variables: \begin{equation}\label{eq:long}
    f(x) = \frac{x^2+3x-x/4}{x}+x^4+ 22/7 x^3 +x^9.
    \end{equation}
    \item A scalar trigonometric function:
    \begin{equation}\label{eq:trig}
    g(x) = \sin(x) + \cos(x) + x^2.
    \end{equation}
    \item A vector-valued polynomial with $s$ variables:
    \begin{equation}\label{eq:poly10}
    h_{s}(x) = 4^s\prod_{i=1}^{s}x_{i}(1 - x_{i}), \qquad  x = (x_{1}, ... , x_{s}).
    \end{equation}
\end{enumerate}

For each expression, we randomly generate evaluation points and plot the average running times over 10 runs. ACORNS generates C code, Mitsuba and Enoki are both running in C++. All the C and C++ code is compiled with \texttt{gcc9}/\texttt{g++9} with the flags \texttt{-O3}, \texttt{-ffast-math}, \texttt{-flto} and, only for C++, \texttt{-std=c++11}.

\begin{figure}

\parbox{0.02\linewidth}{\centering\rotatebox{90}{\scriptsize{Time (s)}}}\hfill\hfill
\parbox{.32\linewidth}{\centering
    Long Polynomial in~\eqref{eq:long}
    \includegraphics[width=\linewidth]{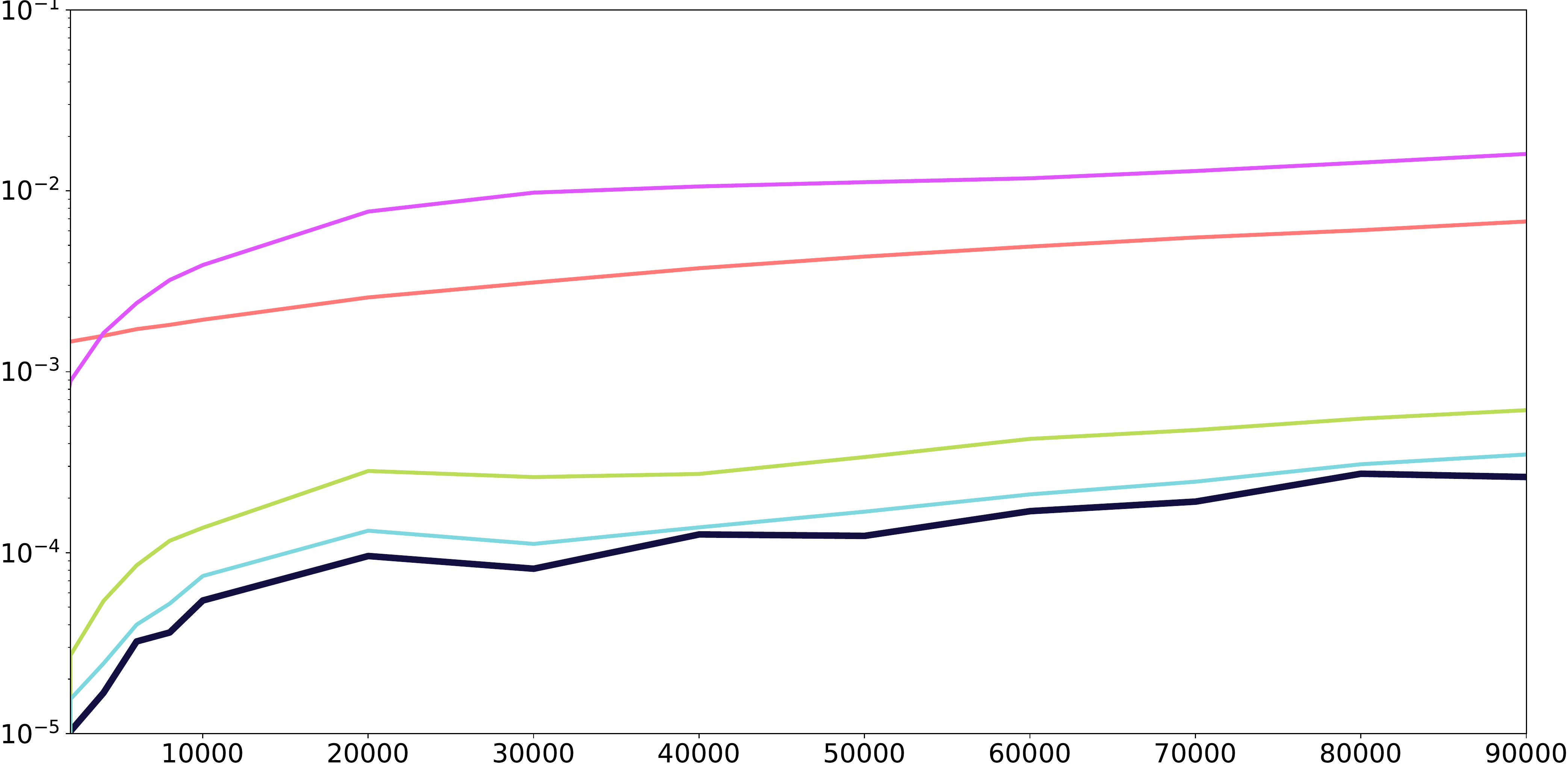}
    \scriptsize{Number of Evaluations}
}\hfill
\parbox{.32\linewidth}{\centering
    Trigonometric Function in~\eqref{eq:trig}
    \includegraphics[width=\linewidth]{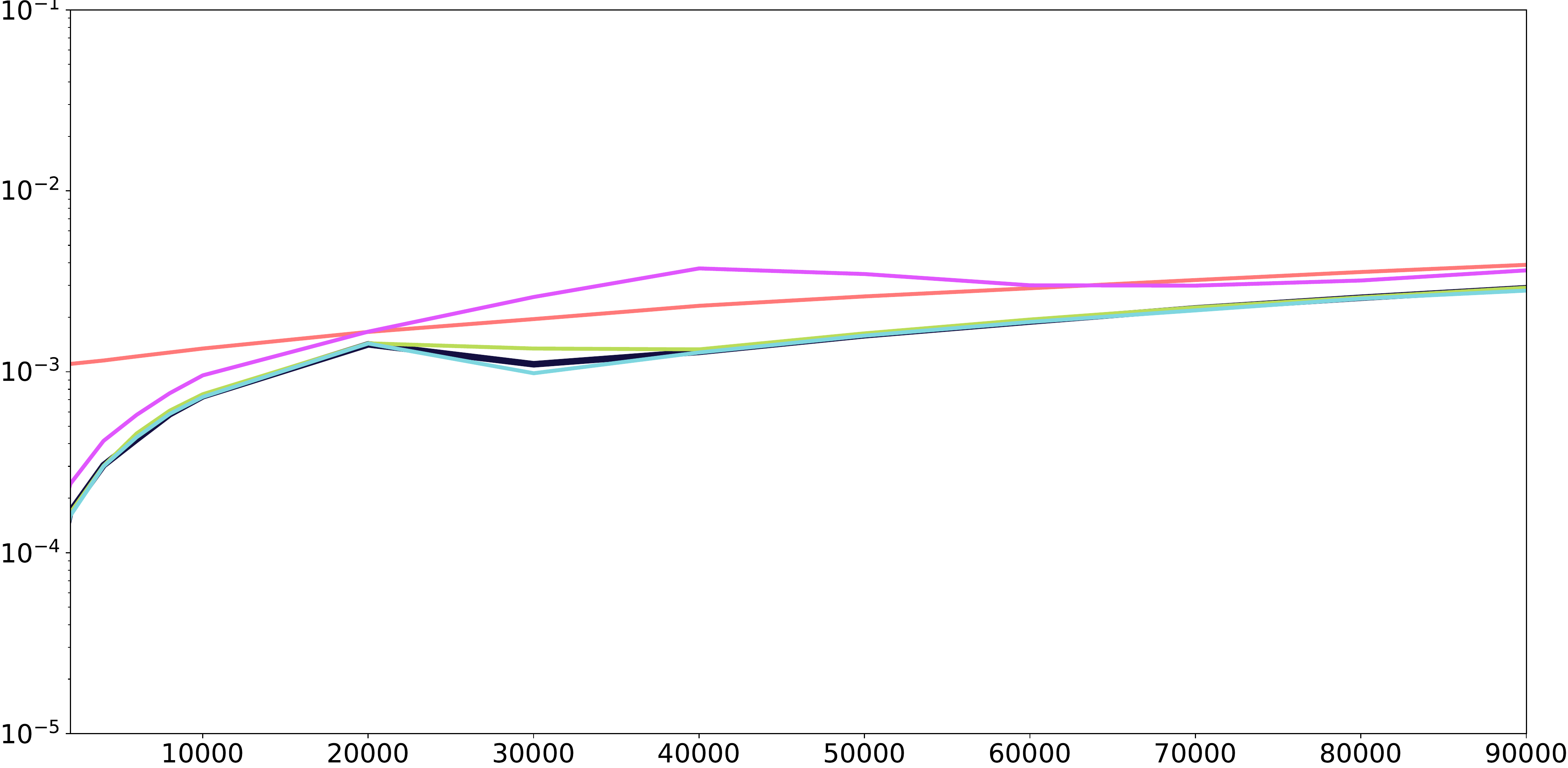}
    \scriptsize{Number of Evaluations}
}\hfill
\parbox{.32\linewidth}{\centering
    Polynomial in~\eqref{eq:poly10} for $s=10$
    \includegraphics[width=\linewidth]{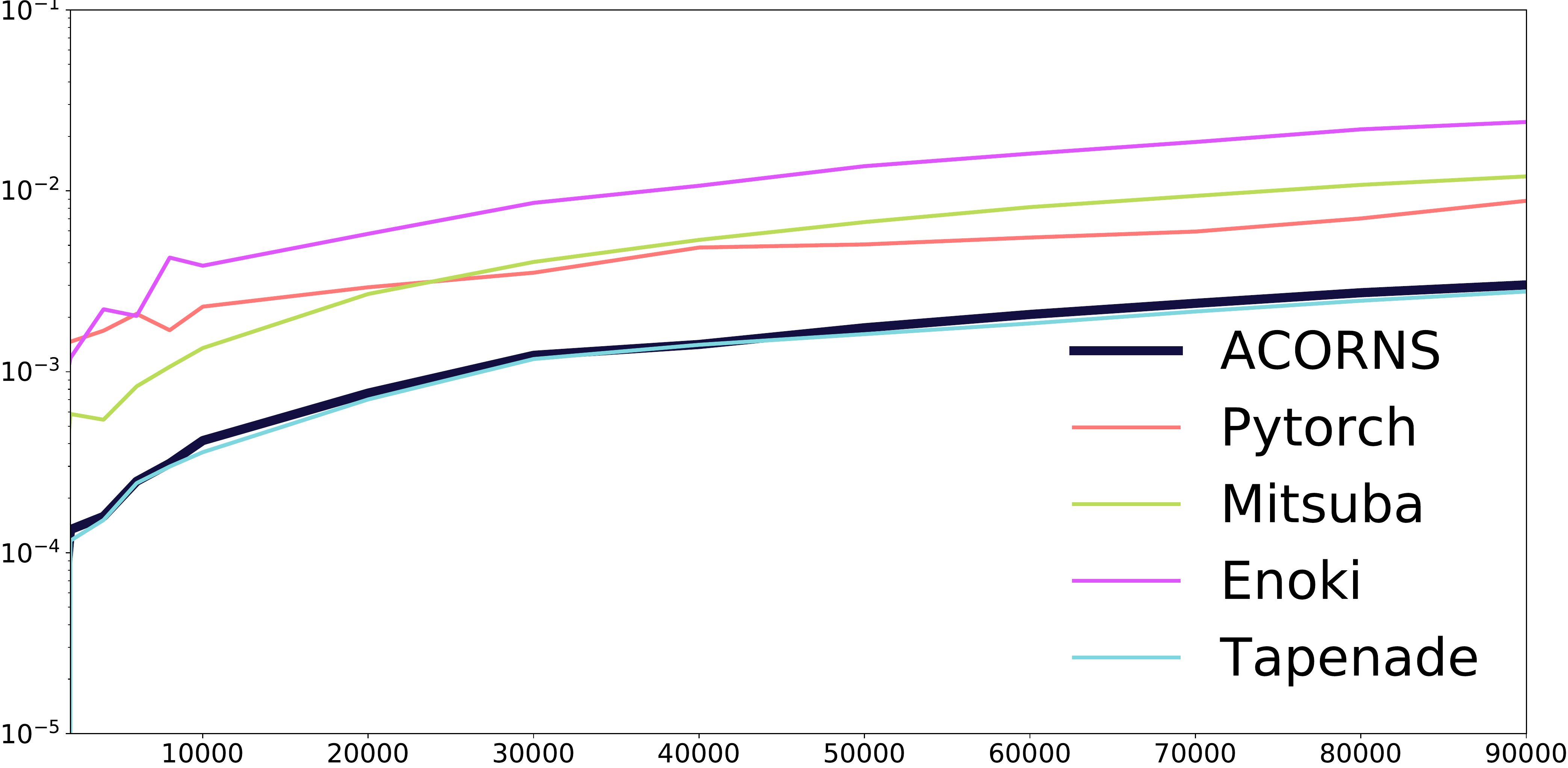}
    \scriptsize{Number of Evaluations}
}
\caption{Average runtimes in log for our different test function for each method.}
\label{fig:comparison-grad}
\end{figure}

\paragraph{Gradient}

In Figure~\ref{fig:comparison-grad} we report the timings, for different number of evaluation points, for the three functions. For the two polynomial functions we are roughly 5$\times$ faster than Mitsuba and Enoki and 2.9 times faster than PyTorch. For the trigonometric function, due to the increase in the cost of evaluating the trigonometric function, our advantage is slightly reduced. For gradients, our approach is very similar to Tapenade, and we obtain comparable timings.

\begin{figure}
\parbox{0.02\linewidth}{\centering\rotatebox{90}{\scriptsize{Time (s)}}}\hfill\hfill
\parbox{.48\linewidth}{\centering
    Gradient
    \includegraphics[width=\linewidth]{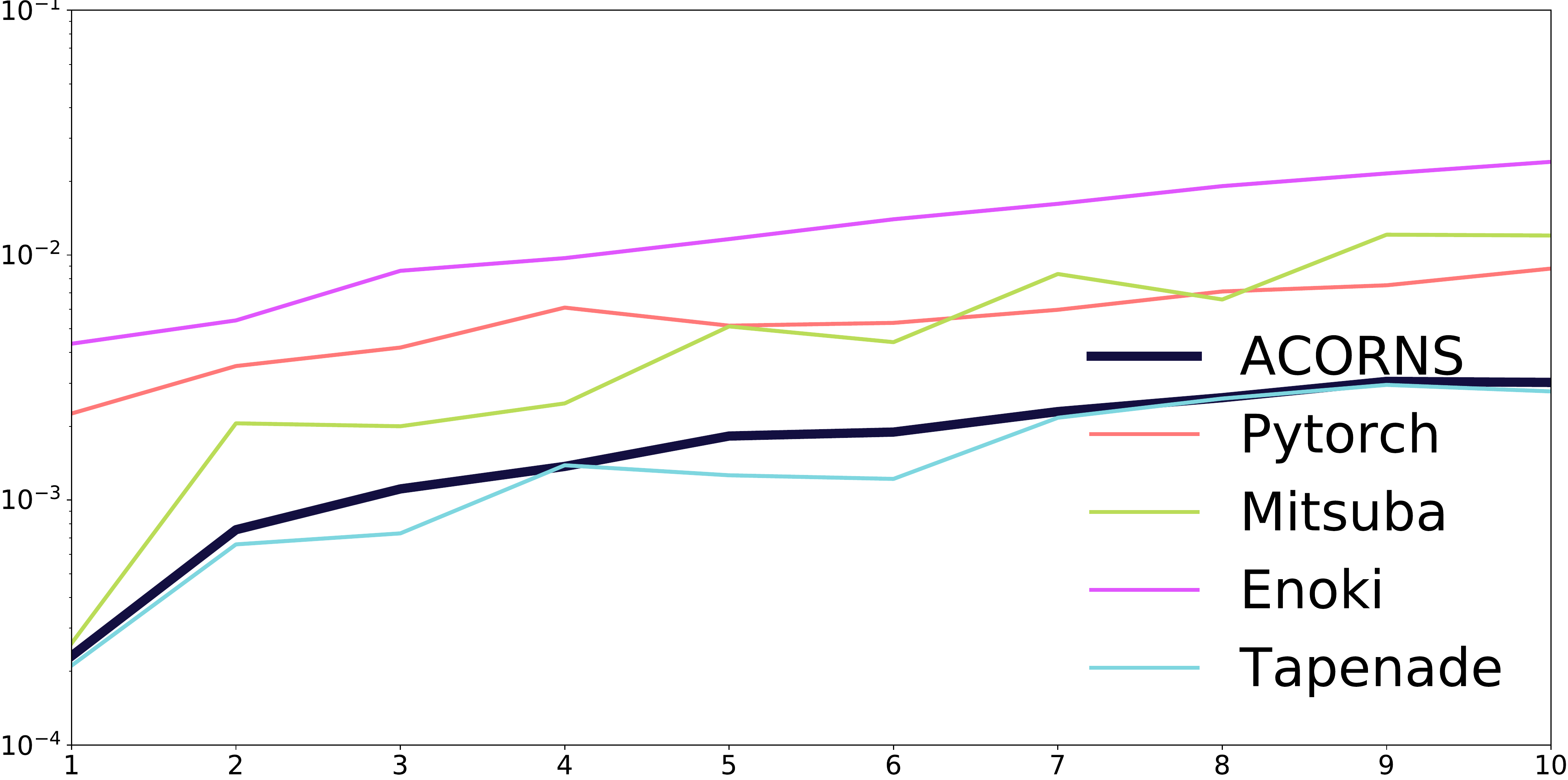}
    \scriptsize{Number of Variables $s$}
}\hfill
\parbox{.48\linewidth}{\centering
    Hessian
    \includegraphics[width=\linewidth]{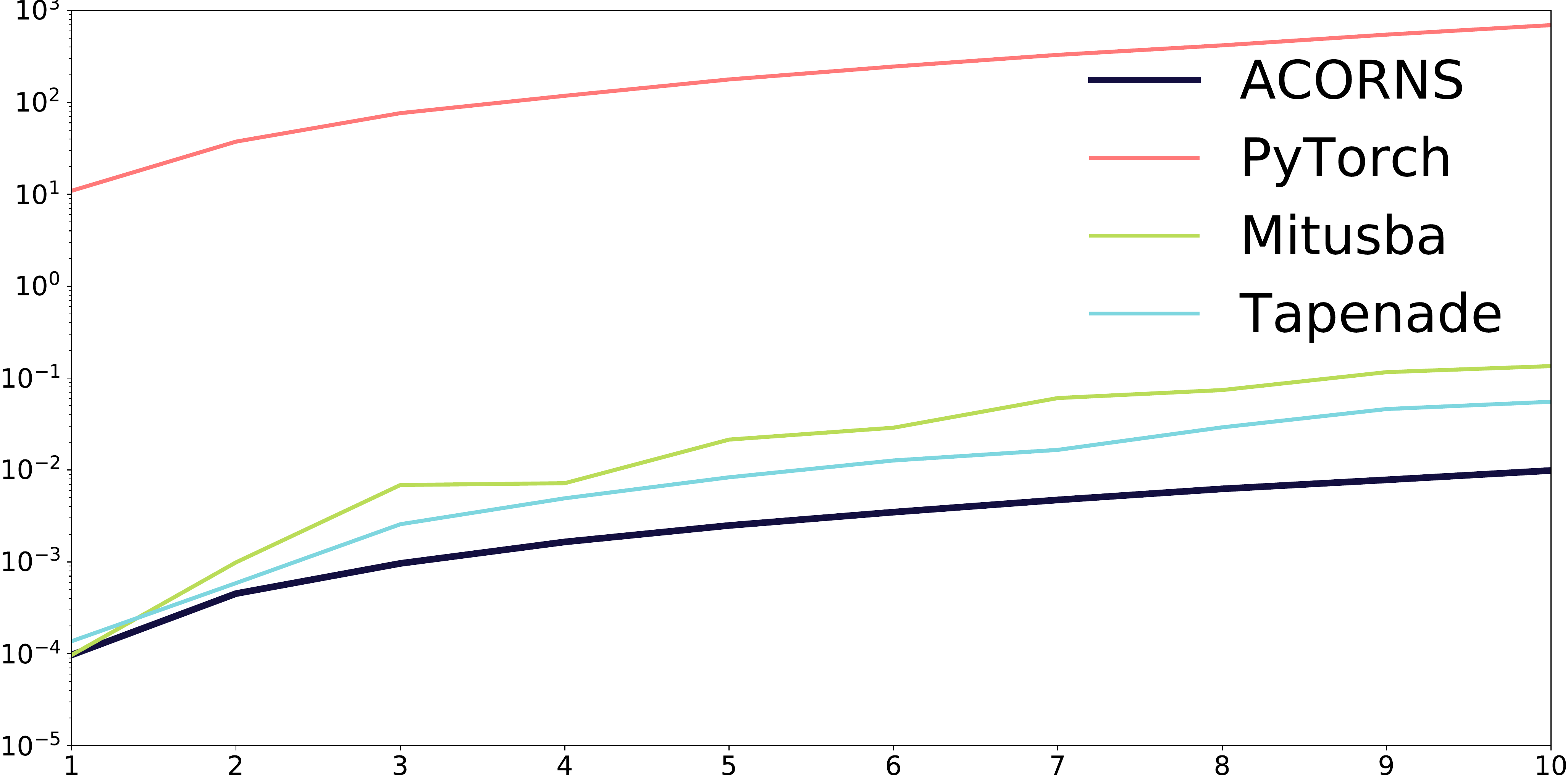}
    \scriptsize{Number of Variables $s$}
}
\caption{Runtimes versus respect to number of variables for the polynomial in~\eqref{eq:poly10} for the different methods.}
\label{fig:comparison-poly-vars}
\end{figure}

In Figure~\ref{fig:comparison-poly-vars} left we fix the number of evaluations and change the parameter $s$ in~\eqref{eq:poly10}, to evaluate how the methods scale with the number of variables. As expected, all methods scale linearly, but our method is still 3.97$\times$  7.95$\times$, and 3$\times$ faster than Mistuba, Enoki, and Pytorch respectively. As for Figure~\ref{fig:comparison-grad}, Tapenade's code has very similar running time as ours.

\paragraph{Hessian}

\begin{figure}
\parbox{0.02\linewidth}{\centering\rotatebox{90}{\scriptsize{Time (s)}}}\hfill\hfill
\parbox{.32\linewidth}{\centering
    Long Polynomial in~\eqref{eq:long}
    \includegraphics[width=\linewidth]{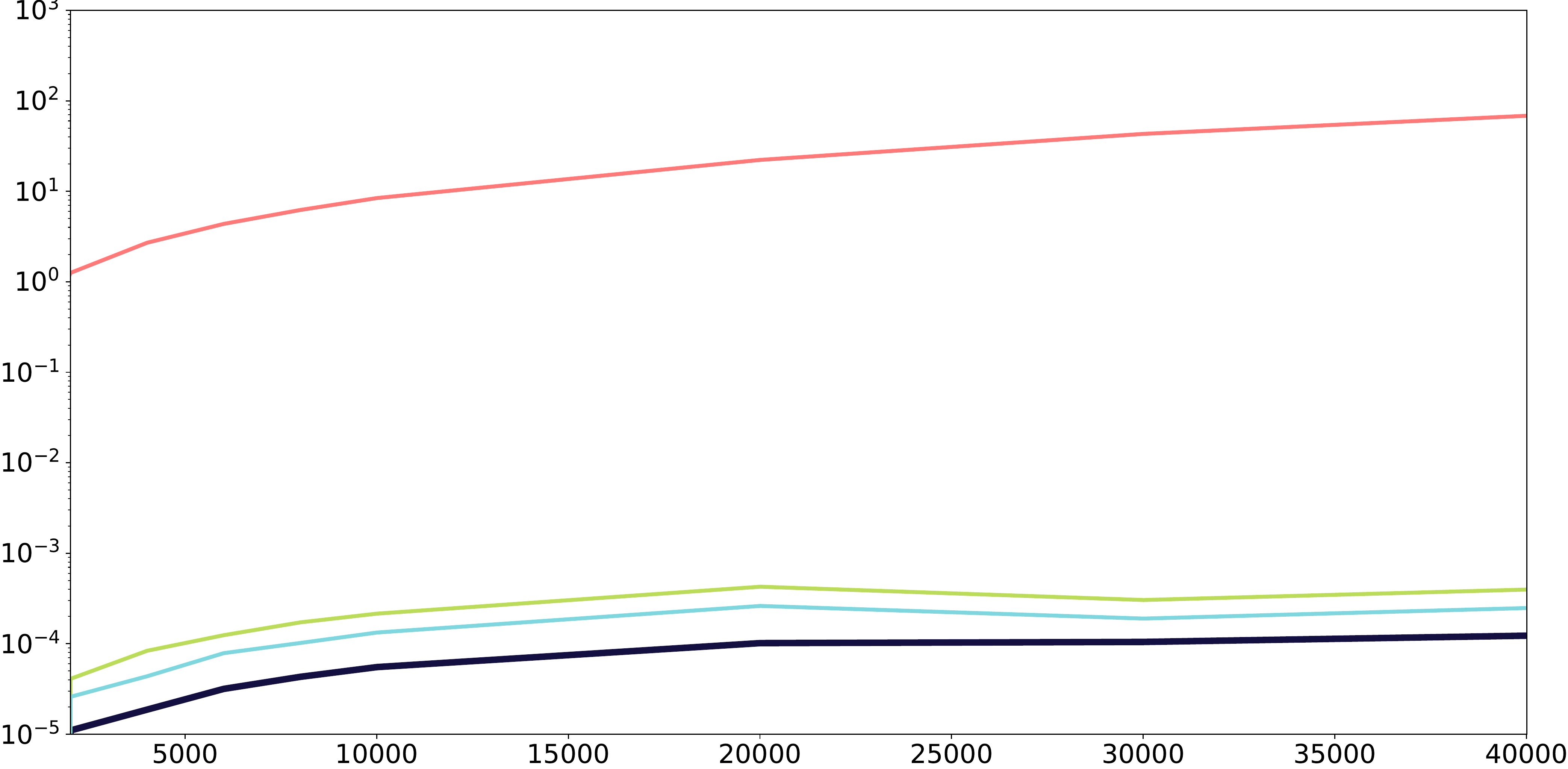}
    \scriptsize{Number of Evaluations}
}
\parbox{.32\linewidth}{\centering
    Trigonometric Function in~\eqref{eq:trig}
    \includegraphics[width=\linewidth]{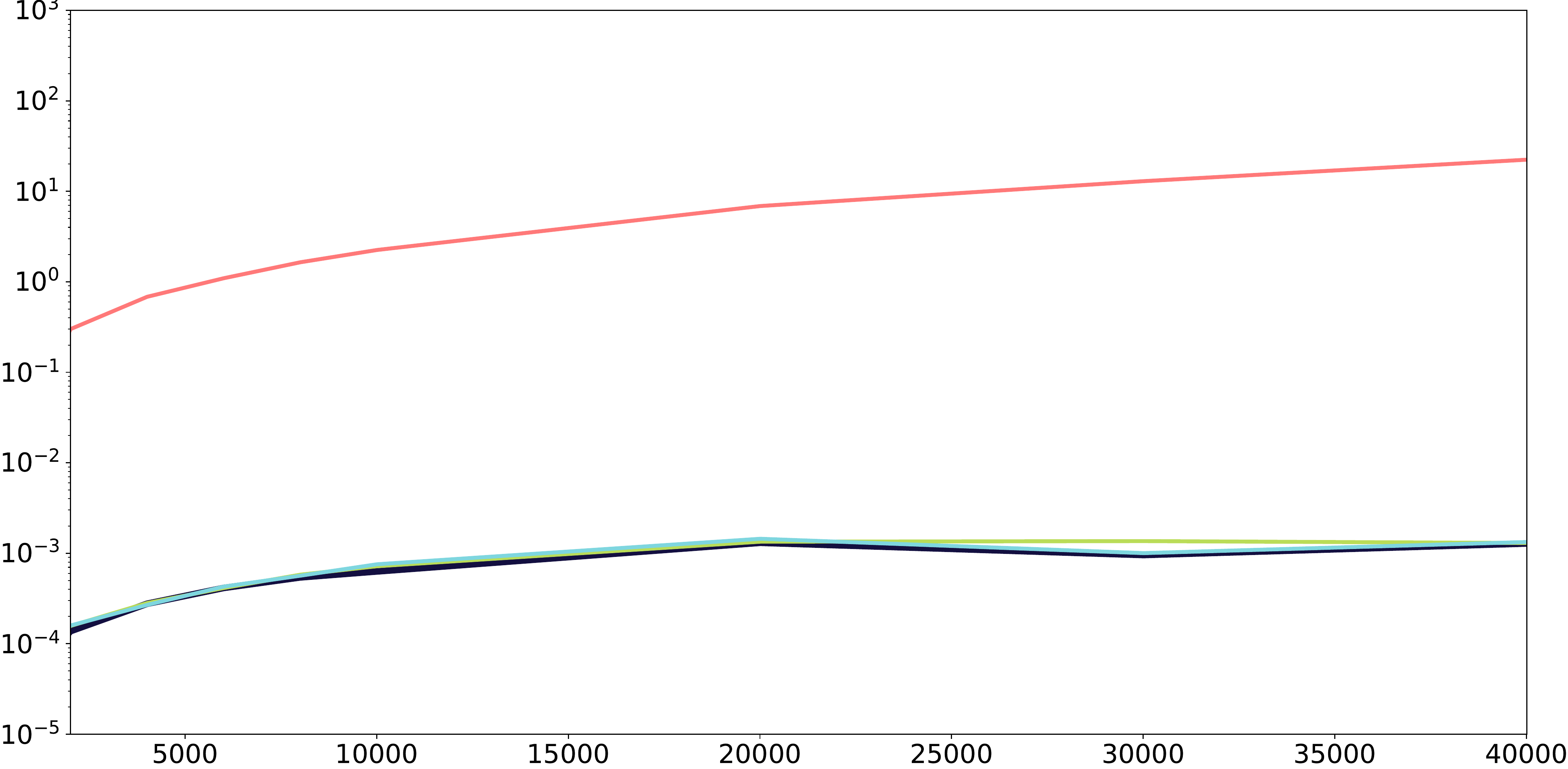}
    \scriptsize{Number of Evaluations}
}
\parbox{.32\linewidth}{\centering
    Polynomial in~\eqref{eq:poly10} for $s=10$
    \includegraphics[width=\linewidth]{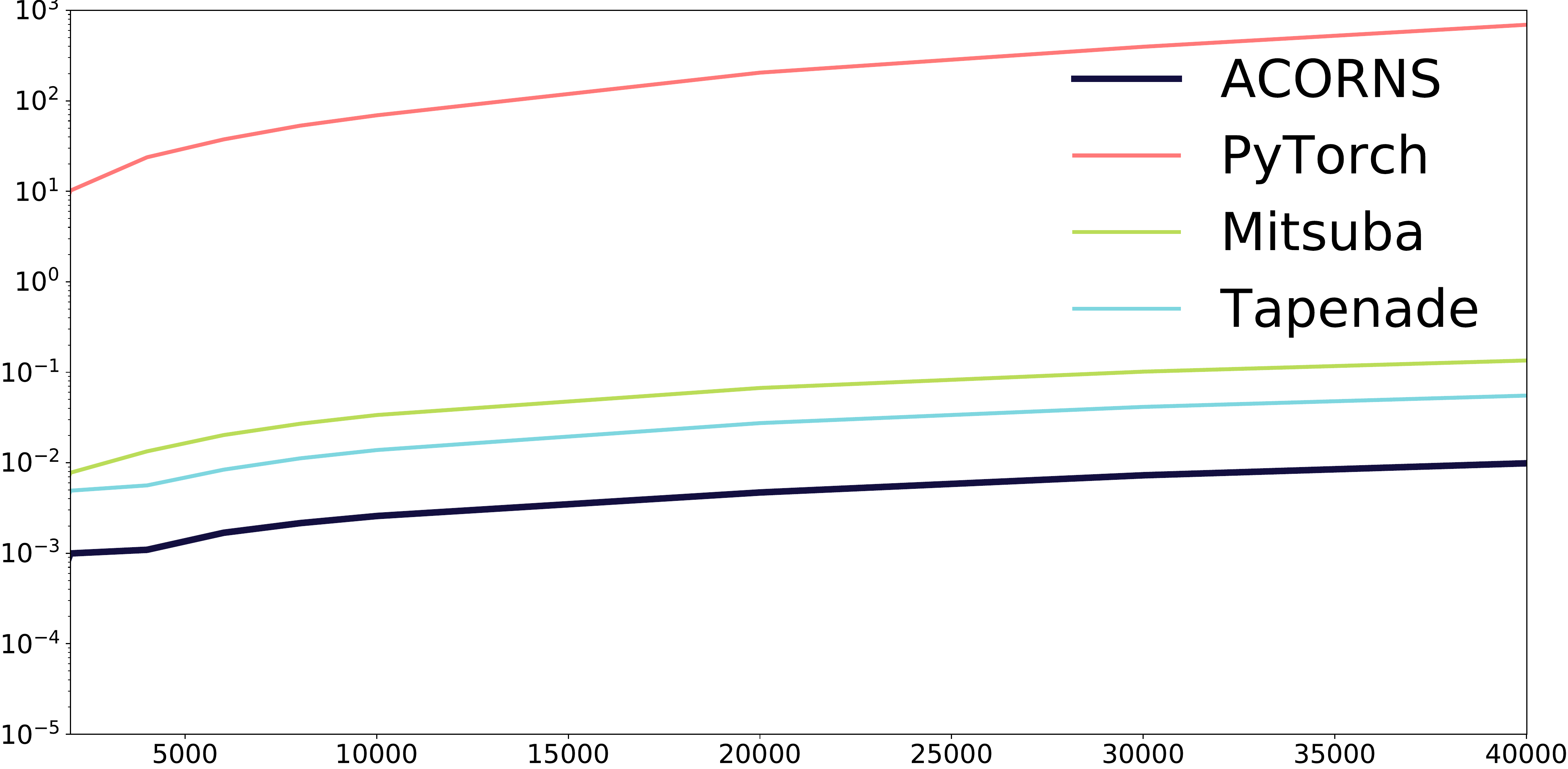}
    \scriptsize{Number of Evaluations}
}
\caption{Runtimes for computing the Hessian of our method for the different test functions.}
\label{fig:comparison-poly-hess}
\end{figure}

 We repeat the same experiments as in previous section to compare ACORNS with Mitsuba, PyTorch, and Tapenade to compute Hessians.  Figure~\ref{fig:comparison-poly-hess} shows that our code is very efficient, $12.8\times$ faster than Mitsuba and $1336\times$ faster than PyTorch (which is not optimized to evaluate Hessians). Tapenade is faster than the other alternatives, but it is still around 5 times slower than our code. Note also that while the Hessian generation is completely automatic with ACORN, it does require manual interaction with Tapenade, since Hessian computation is not fully supported.
 
 The scaling with respect to the number of variables for Hessian evaluation (Figure~\ref{fig:comparison-poly-vars} right) has a similar trend as for the gradients.

% \begin{figure}
% \parbox{0.02\linewidth}{\centering\rotatebox{90}{\scriptsize{Time (s)}}}\hfill\hfill
% \parbox{.95\linewidth}{\centering
%     \includegraphics[width=\linewidth]{graphs/graph_max_hess_g++9.pdf}
%     \scriptsize{Number of Variables $s$}
% }
% \caption{Runtimes versus respect to number of variables for the polynomial in~\eqref{eq:poly10} for the different methods \DP{merge with below, larger fonts}}
% \label{fig:comparison-poly-vars}
% \end{figure}

% \TS{i would drop the table, and the other plot, i think it is repeated.}

% When we compare the runtimes of this very complex algorithmic function against Mitsuba (as it's the only one that handles Hessians) we see in Figure \ref{fig:speedup-mitsuba} that we do much better than Mitsuba on this as well. Looking at Table \ref{tab:mitsuba-comparison}. shows precisely how much better we do than Mitsuba for Hessians of various variable sizes.

% \begin{table}
% \centering
%  \begin{tabular}{||c c c||} 
%  \hline
%  Vars & Us vs. Mitsuba Static & Us vs. Mitsuba Dynamic \\ [0.5ex] 
%  \hline\hline
%  12 & 3.84x & 10.55x \\ 
%  30 & 12.92x & 19.52x \\
%  60 & - & 10.92x \\
%  105 & - & 6.72x \\ [1ex] 
%  \hline
%  \end{tabular}
%  \caption{Speedup of Us vs. Mitsuba}
%  \label{tab:mitsuba-comparison}
% \end{table}

% DISCUSS:
% \begin{figure}
% \includegraphics[width=1\linewidth]{graphs/us_vs_wenzel_raw.pdf}
% \caption{Speedup of us over Mitsuba for algorithmic function with respect to number of variables}
% \label{fig:speedup-mitsuba}
% \end{figure}

\paragraph{Parallel Evaluation}

\begin{figure}
\parbox{0.02\linewidth}{\centering\rotatebox{90}{\scriptsize{Speedup}}}\hfill
\parbox{.45\linewidth}{\centering
    \includegraphics[width=\linewidth]{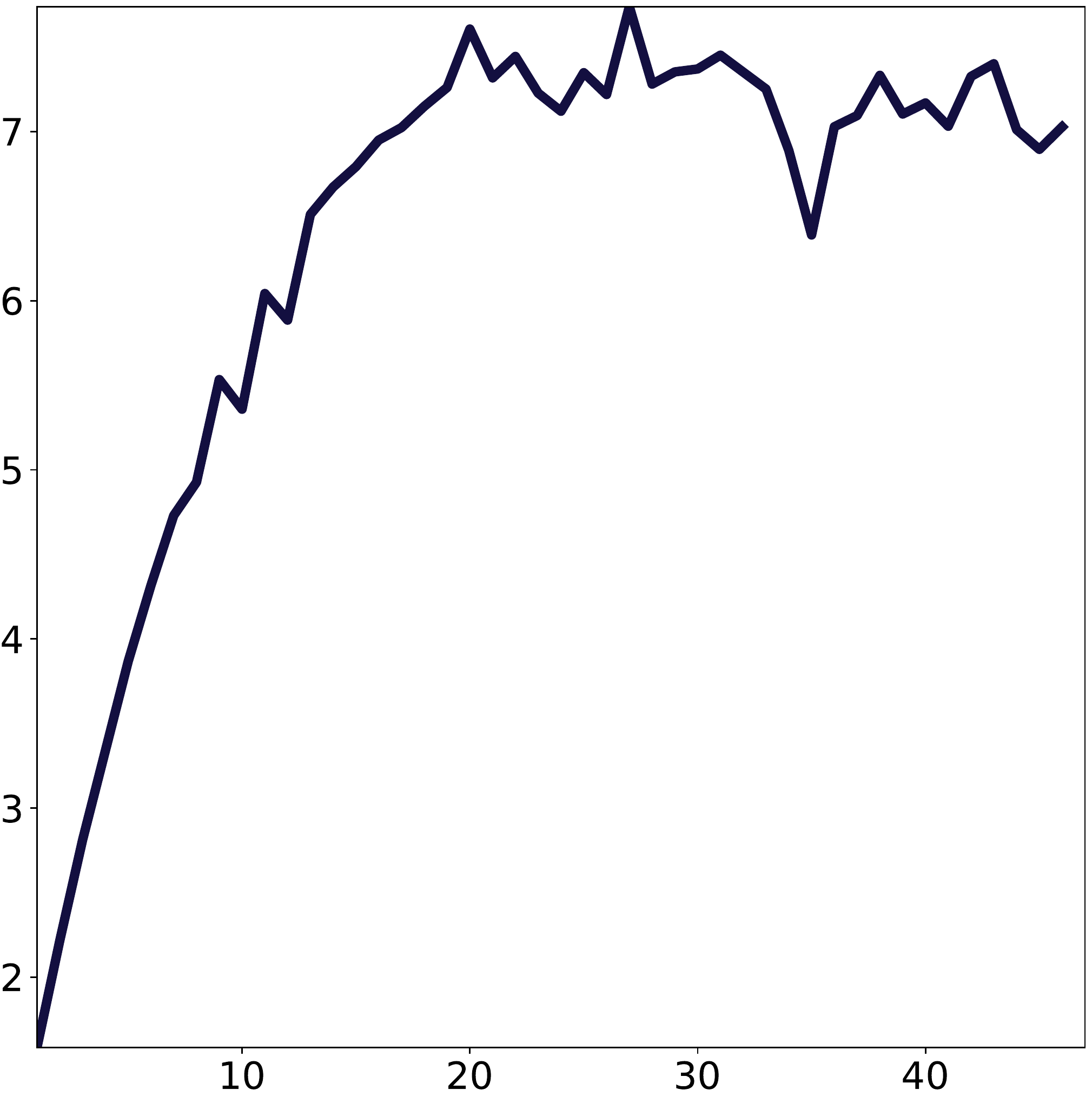}
    \scriptsize{Number of Threads}
}
\parbox{0.02\linewidth}{\centering\rotatebox{90}{\scriptsize{Time}}}\hfill
\parbox{.47\linewidth}{\centering
    \includegraphics[width=\linewidth]{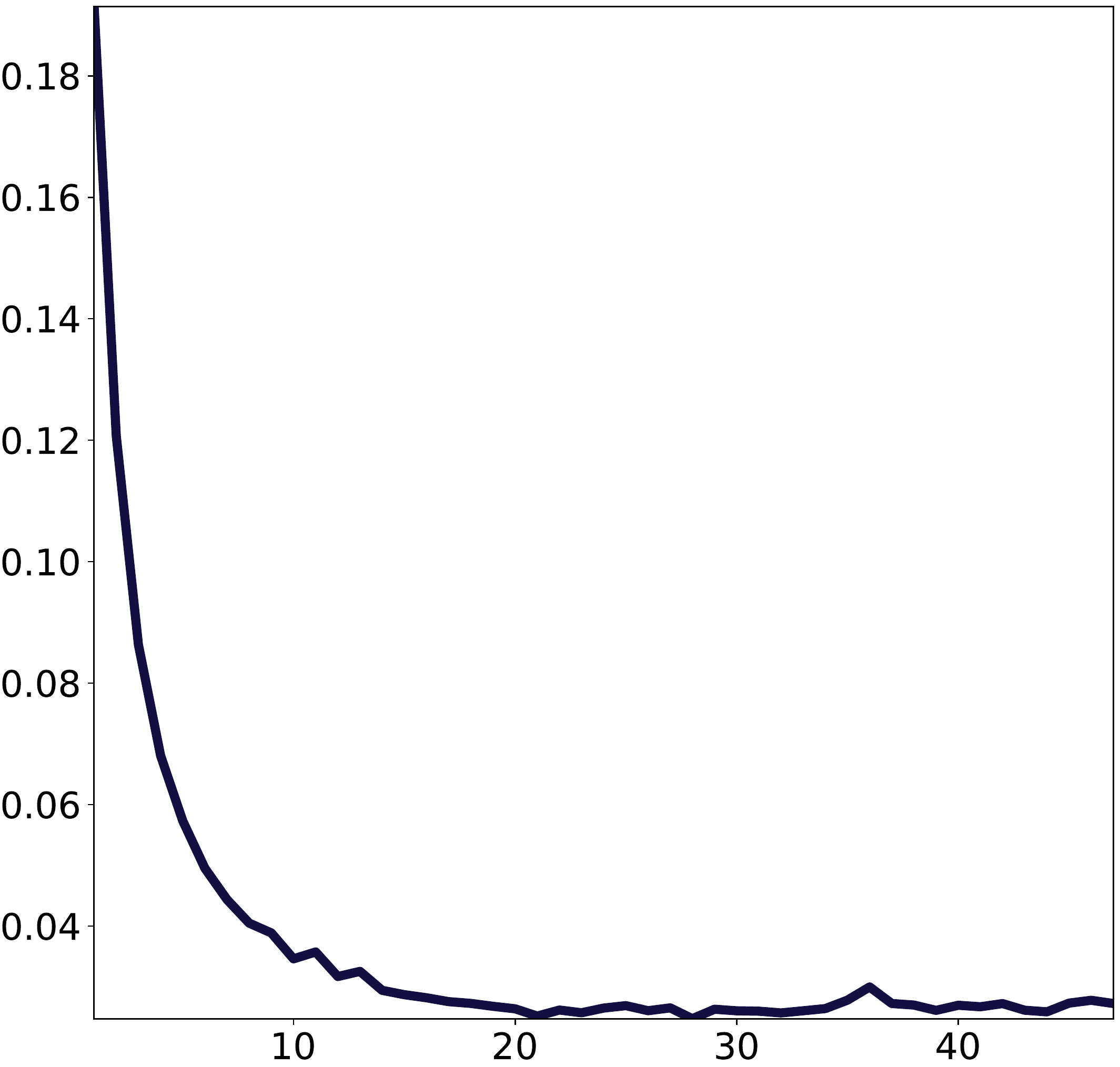}
    \scriptsize{Number of Threads}
}
%%%%%%%%%%%%%%
% \hfill\hfill\hfill
%%%%%%%%%%%%
% \parbox{0.02\linewidth}{\centering\rotatebox{90}{\scriptsize{Time (s)}}}\hfill
% \parbox{.45\linewidth}{\centering
%     \includegraphics[width=\linewidth]{graphs/parallel-graph.pdf}
%     \scriptsize{Number of Threads}
% }
\caption{Time and speedup of our method over number of threads for computing the Hessian expression of $h_{25}$~\eqref{eq:poly10} for 100000 points.}
\label{fig:hess-scalability}
\end{figure}

We ran the function in \eqref{eq:poly10} for $s=25$ and computed the gradient for 100000 evaluation points allowing OpenMP to use a different number of threads. We observe an almost linear scaling up to 15 threads (Figure~\ref{fig:hess-scalability}, right) on our workstation, and it then flattens as more threads are added. We speculate that this is due to saturation of the memory bandwidth of the workstation, since our task is inherently massively parallelizable.

\section{Applications}

We integrated ACORNS in two applications in geometry processing and in physical simulation to evaluate the  gain in performance in a realistic setting. In both these cases, the algorithm needs to assemble a dense gradient and a sparse Hessian. For both applications, ACORNS is used to compute small dense blocks, which are then assembled in a larger sparse Hessian matrix. Directly supporting the automatic construction of the sparse Hessian is an interesting avenue for future work.

\label{sec:applications}
\subsection{Parametrization}
Given a surface embedded in three dimensional space, an important problem in geometry processing is to assign parametric values to it, i.e. computing a flattening of the surface on a plane. We refer the readers to  \cite{floater2005surface,sheffer2007mesh} for a comprehensive survey.
In our application, we consider minimizing the non-convex symmetric Dirichlet energy \cite{smith2015bijective} with projected Newton's method \cite{teran2005robust}: when the Hessian is not positive (semi-)definite, possibly at saddle points, we project it to the closest positive definite matrix through polar decomposition.
The algorithm involves computing the full Jacobians and Hessians per element: our original implementation uses Mitsuba, and we replaced it with ACORNS. We summarize the statistics in Table~\ref{tab:tab-hilbert}. The autodiff time is reduce considerably, but in this application is not the bottleneck, thus resulting in a small reduction of the overall running time.

\begin{table}
    \centering
    \parbox{.5\linewidth}{\centering
    Input
    \includegraphics[width=\linewidth]{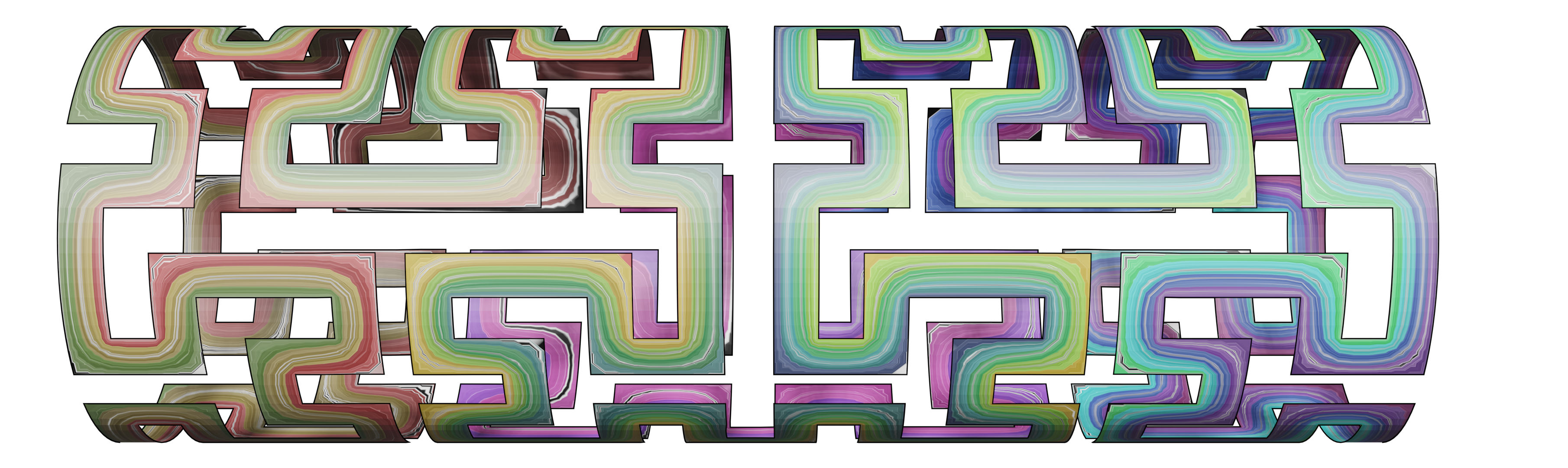}
    % }\hfill
    % \parbox{.6\linewidth}{\centering
    Output
    \includegraphics[width=\linewidth]{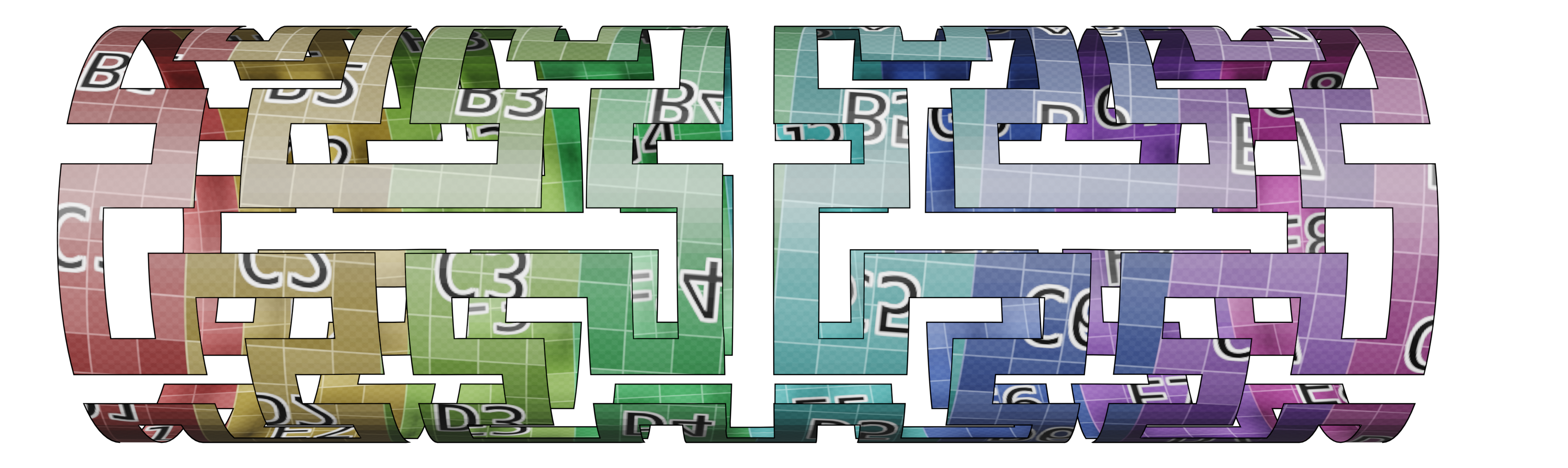}
    }%\hfill\hfill
    \parbox{.5\linewidth}{\begin{tabular}{ccc|cc}
    \multicolumn{3}{c}{Autodiff Time}& \multicolumn{2}{|c}{Total Time}\\
    Ours&Mitsuba&Ratio&    Ours&Mitsuba\\
    \hline
    0.23s&0.40s&1.75 % AD 
    &3.66s&4.27s\\ % Total 
    \end{tabular}}
    \caption{Optimize the parameterization of a surface mesh.}
    \label{tab:tab-hilbert}
\end{table}

\subsection{Finite Element Simulation}
The finite element method aims at solving partial differential equations (PDEs) describing physical phenomena. A typical example is to simulate the deformation of an elastic body subject to some external forces (boundary conditions); the algorithm involves minimizing an elastic energy, and requires computing derivatives. %While for simple linear modelen the boundary conditions induce a large deformation, a simple linear model is not descriptive enough and one usually relies on non-linear elastic energy such as Neo-Hookean, which is solved using the Newton method. 
We integrate ACORNS in PolyFEM~\cite{polyfem} and compared, for different finite element discretization order (ranging from linear to cubic), the timings of a Neo-Hookean simulation using the Mitsuba autodiff and with ACORNS to compute Hessian and gradient of the energy. In the finite element method, all computations are performed locally on each tetrahedron. The different order will have different local number of degree of freedom (which determine the size of the gradient and Hessian): a linear tetrahedron has 12, a quadratic has 30, and a cubic has 60. Our method provides an overall speedup of 2 to 4 times, depending on the degree, over the entire simulation time (Table~\ref{tab:tab-polyfem}).

\begin{table}
    \centering
    \parbox{.1\linewidth}{\centering
    Input
    \includegraphics[width=\linewidth]{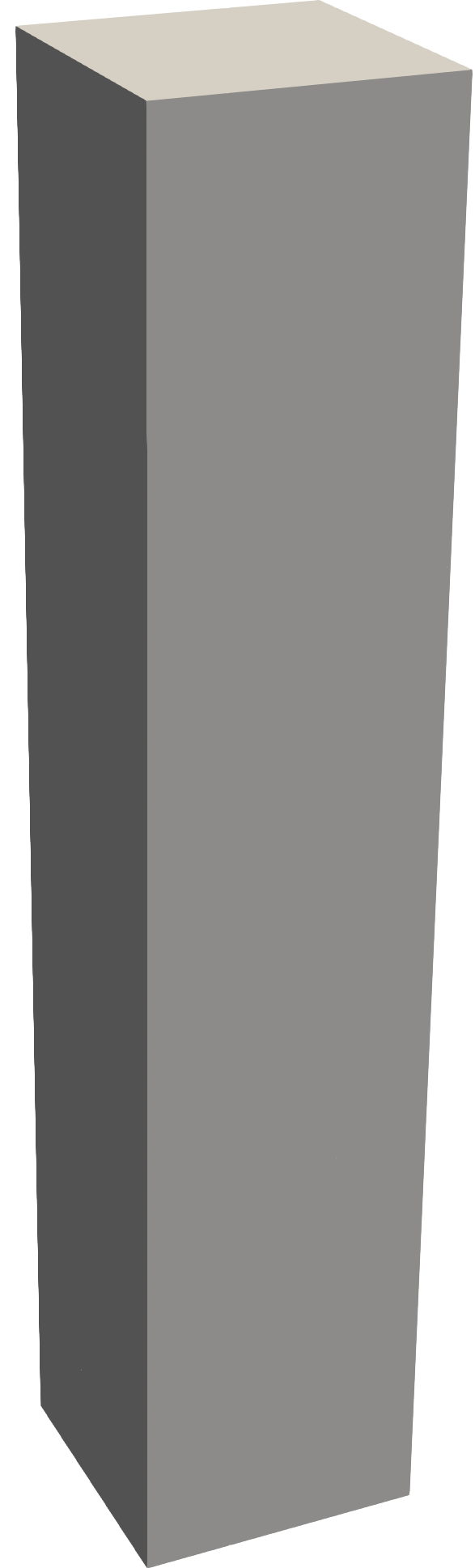}
    }\hfill
    \parbox{.1\linewidth}{\centering
    Output
    \includegraphics[width=\linewidth]{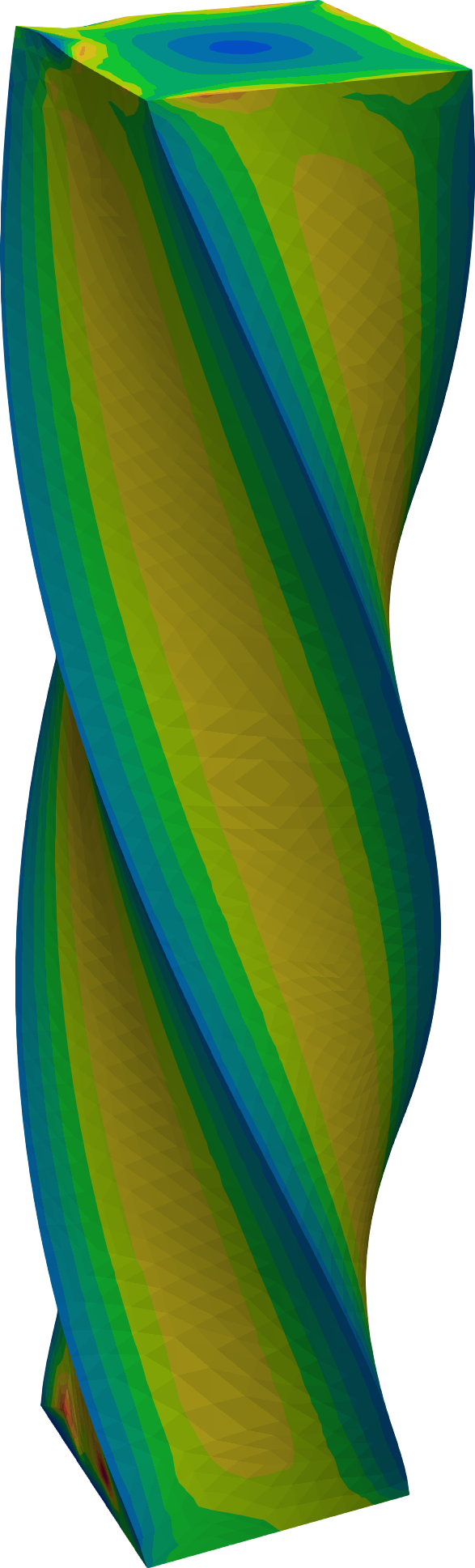}
    }\hfill\hfill
    \parbox{.7\linewidth}{
    \begin{tabular}{c|ccc|ccc}
    &       \multicolumn{3}{c}{Hessian Time}& \multicolumn{3}{|c}{Total Time}\\
    Degree& Ours&Mitsuba&Ratio&    Ours&Mitsuba&Ratio\\
    \hline
    1       &0.14s&0.98s&7.1         &12.43s&55.92s&4.5\\
    2       &1.18s&5.91s&5.0         &110.36s&310.61s&2.8\\
    3       &8.14s&34.84s&4.2        &955.71s&2076.08s&2.2
    \end{tabular}
    }
    \caption{Runtime of a Neo-Hookean elastic simulation for different FE degrees.}
    \label{tab:tab-polyfem}
\end{table}
% --------------
% P3: 
% assembly time 8.13695s
% assembly time 34.8445s
% 4.2x

% [2020-05-28 19:16:50.017] [polyfem] [info]  took 955.705s vs
% [2020-05-28 19:37:41.039] [polyfem] [info]  took 2076.08s
% 2.2x

% --------------
% P2:
% assembly time 1.18022s
% assembly time 5.9123s
% 5.0x

% [2020-05-29 09:51:33.397] [polyfem] [info]  took 110.355s
% [2020-05-29 09:55:32.379] [polyfem] [info]  took 310.614s
% 2.8x
% --------------
% P1:
% assembly time 0.138068s
% assembly time 0.978083s
% 7.1x

% [2020-05-29 09:52:14.561] [polyfem] [info]  took 12.4343s
% [2020-05-29 09:54:06.197] [polyfem] [info]  took 55.92s
% 4.5x

\section{Concluding Remarks}

We introduced ACORNS, a software library for automatic differentiation that generates efficient kernels for computing both gradients and Hessian and can be easily integrated in existing C or C++ projects. The core idea of our algorithm is to fully unroll the computational graph, and rely on modern compilers to optimize the resulting code. Compared with alternatives, our algorithm is faster at evaluation time, but slower during compilation: we believe that this tradeoff is very interesting for many scientific computing applications where the evaluation of small dense Hessian is repeated millions of times.
The compilation time could be reduced by performing symbolic simplification on the code before generating the c files. In addition, we could avoid the parsing time of the compiler by directly generating bytecode (for example for the llvm compiler) and compiling it in memory, avoiding unnecessary disk access.
Variable conditional are not support by ACORNS. Support for it could be added, but the performance will likely decrease, due to the reduced optimization options and more complex parallelization.

We would like to extend ACORNS to directly support the generation of sparse Hessians \cite{gebremedhin2009efficient,walther2008computing} and also extend the code generation to target GPU accelerators. 

Our open-source reference implementation and a set of examples on how to use it is available at https://github.com/deshanadesai/acorns, and it is also released as a Conda package on conda-forge to allow an easy installation on all major operating systems.

% \bibliographystyle{ACM-Reference-Format}
% \bibliography{references}

\appendix

\end{document}